\documentclass[journal]{IEEEtran}
\usepackage[T1]{fontenc}
\usepackage{xspace,amsmath,amssymb,amsthm,amsfonts,epsfig,syntonly,dsfont}
\usepackage{graphicx,empheq}
\usepackage{xcolor,bm}

\usepackage{mathtools}
\usepackage{algpseudocode,algorithm}
\usepackage{setspace}
\usepackage{balance}
\usepackage{url}

\DeclareMathOperator*{\argmin}{arg\,min}
\definecolor{mymage}{RGB}{218, 47, 123}
\definecolor{amber}{rgb}{1.0, 0.49, 0.0}
\definecolor{fuchsia}{rgb}{0.57, 0.36, 0.51}
\definecolor{rev}{rgb}{0.1, 0.4, 0.99}

\theoremstyle{definition}\newtheorem{remark}{Remark}
\usepackage{bbm}
\usepackage{amssymb}
\usepackage{setspace}
\usepackage{wrapfig}
\usepackage{times,color}
\usepackage{caption}
\usepackage{subcaption}
\usepackage{algorithm}
\usepackage{algpseudocode}
\usepackage{epstopdf}
\usepackage{cite,footnote,xspace,syntonly,bm}

\input{mysymbol.sty}
\allowdisplaybreaks[4]

\algnewcommand{\Inputs}[1]{%
  \State \textbf{Inputs:}
  \Statex \hspace*{\algorithmicindent}\parbox[t]{.8\linewidth}{\raggedright #1}
}
\algnewcommand{\Initialize}[1]{%
  \State \textbf{Initialize:}
  \Statex \hspace*{\algorithmicindent}\parbox[t]{.8\linewidth}{\raggedright #1}
}

\def \bbeta {{\boldsymbol \beta}}

\def \bbzeta {{\boldsymbol \zeta}}

 \long\def\symbolfootnote[#1]#2{\begingroup
 	\def\thefootnote{\fnsymbol{footnote}}
 	\footnote[#1]{#2}\endgroup} \psfull

\DeclareMathOperator*{\opt}{optimize}
\DeclareMathOperator*{\mini}{minimize}
\DeclareMathOperator*{\maxi}{maximize}
\DeclareMathOperator*{\st}{subject~to}

\begin{document}

\title{Learning and Management for Internet-of-Things:\\ Accounting for Adaptivity and Scalability}
\author{
Tianyi Chen, Sergio Barbarossa,  Xin Wang, Georgios B. Giannakis, and Zhi-Li Zhang
\vspace{-0.5cm}
\thanks {The work of T. Chen and G. B. Giannakis was supported by NSF 1509040, 1508993, and 1711471. 
The work of S. Barbarossa was supported by the H2020 EUJ Project 5G-MiEdge, Nr. 723171.
The work of X. Wang was supported by the National Natural Science Foundation of China Grants No. 61671154, the National Key Research and Development Program of China Grant 2017YFB0403402, and the Innovation Program of Shanghai Municipal Science and Technology Commission 17510710400. 
The work of Z.-L. Zhang was supported by US DoD HDTRA1-14-1-0040, NSF 1411636 and 1617729.}
\thanks{T. Chen and G. B. Giannakis are with the Digital Technology Center, University of Minnesota, Minneapolis, MN 55455 USA. Emails: \{chen3827, georgios\}@umn.edu. S. Barbarossa is with the Department of Information Engineering, Electronics, and Telecommunications, Sapienza University of Rome, 00184, Rome, Italy. E-mail: sergio.barbarossa@uniroma1.it. X. Wang is with the Shanghai Institute for
Advanced Communication and Data Science, the Key Laboratory for Information Science of
Electromagnetic Waves (MoE), Department of Communication Science and
Engineering, Fudan University, Shanghai 200433, China. E-mail: xwang11@fudan.edu.cn.
Z.-L. Zhang is with Department of Computer Science, University of Minnesota, Minneapolis, MN 55455 USA. E-mail: zhang@cs.umn.edu.
}
}

\markboth{}{}

\maketitle

\begin{abstract}
Internet-of-Things (IoT) envisions an intelligent infrastructure of networked smart devices offering task-specific monitoring and control services. The unique features of IoT include extreme heterogeneity, massive number of devices, and unpredictable dynamics partially due to human interaction. These call for foundational innovations in network design and management. Ideally, it should allow efficient adaptation to changing environments, {and low-cost implementation scalable to massive number of devices}, subject to stringent latency constraints. 
To this end, the overarching goal of this paper is to outline a unified framework for online learning and management policies in IoT through joint advances in communication, networking, learning, and optimization. 
From the network architecture vantage point, the unified framework leverages a promising fog architecture that enables smart devices to have proximity access to cloud functionalities at the network edge, along the cloud-to-things continuum. 
From the algorithmic perspective, key innovations target online approaches adaptive to different degrees of nonstationarity in IoT dynamics, and their {scalable model-free implementation} under limited feedback that motivates blind or bandit approaches. The proposed framework aspires to offer a stepping stone that leads to systematic designs and analysis of task-specific learning and management schemes for IoT, along with a host of new research directions to build on.
\end{abstract}

\section{Introduction}
\label{sec:intro}

\begin{wrapfigure}{R}{0.245\textwidth}
\vspace{-1mm}
\hspace{1mm}
	\includegraphics[scale=0.5]{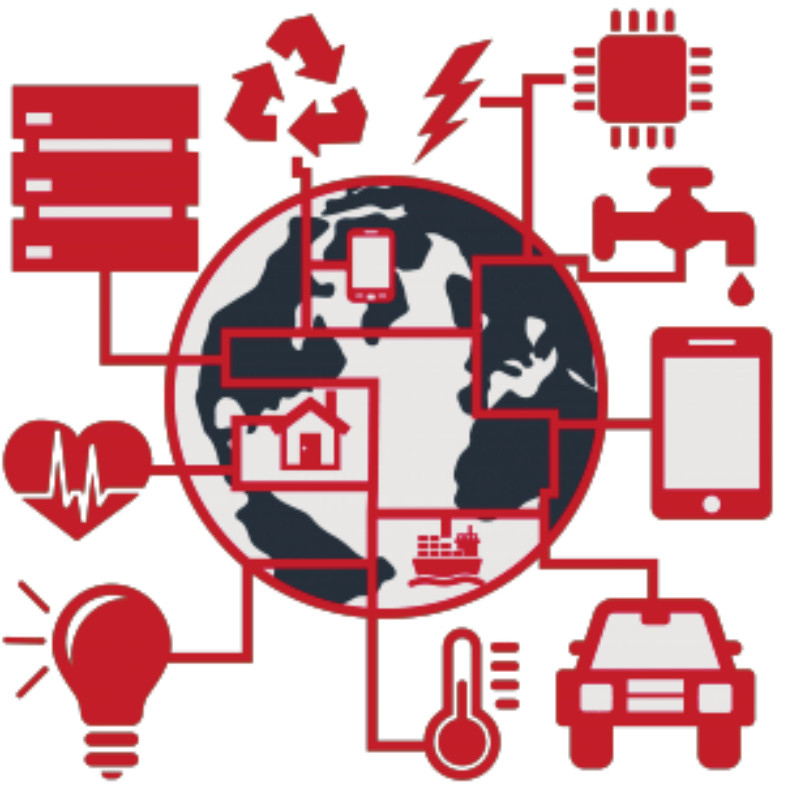}
	\vspace{-1mm}\caption{\hspace{-1mm}{Internet of Everything\! \cite{ioeweb}.\!}}
	\label{fig:IoE}
	\vspace{-1mm}
\end{wrapfigure}

The past decade has witnessed a proliferation of connected devices and objects, where the notion of  Internet-of-Things (IoT) plays a central role in the envisioned technological advances.  Conceptually speaking, IoT foresees an intelligent network infrastructure with ubiquitous smart devices - home automation, interactive healthcare, and self-driving connected vehicles, are typical in IoT \cite{atzori2010,zanella2014internet}; see Fig. \ref{fig:IoE}.
Today, a number of IoT applications have already brought major benefits to many aspects of our daily life. 
The current generation of IoT can already afford an increasing amount of real-time automation, and thus intelligence toward the vision of real-time IoT. 
However, despite the popularity of IoT, several critical challenges must be addressed before embracing its full potential \cite{stankovic2014,al2015internet}.
To this end, we highlight three key challenges that are arguably expected to be at the epicenter of emerging IoT research fields.

\vspace{0.1cm}
\noindent\textbf{Extreme heterogeneity}.
The computational and communication capacities of connected devices differ due to differences in hardware (e.g., CPU frequency), communication protocol (e.g., ZigBee, WiFi), and energy availability (e.g., battery level) \cite{zachariah2015}.
The tasks carried out on various devices are often considerably diverse, e.g., motion sensors monitor human behavior in a smart home \cite{lymberopoulos2011}, while cameras are responsible for recognizing a suspicious behavior in a crowded environment, or, vehicle plates in a parking garage.

\vspace{0.1cm}
\noindent\textbf{Unpredictable dynamics}.
Unlike many existing communication, computing and networking platforms, the IoT dynamics can stem from multiple sources, where \emph{adaptivity} is not only critical but also essential in designing hardware and management protocols. Such sources entail human-in-the-loop dynamics in addition to physical objects \cite{lymberopoulos2011}, demand response in energy systems \cite{giannakis2013}, and intelligent automotive applications \cite{lu2014}. In these applications, IoT dynamics are intertwined with or even partially determined by human behavior \cite{munir2013,nunes2015,duan2017} - as such, high degree of adaptivity in the  algorithm and hardware design is needed.

\vspace{0.1cm}
\noindent\textbf{Scalability at the core}.
IoT entails an intelligent network infrastructure with a massive number of devices. It is estimated that by 2020, there will be more than 50 billion devices connected through the Internet \cite{gerhardt2012}, which highlights \emph{scalability} as a key challenge for IoT \cite{stankovic2014,atzori2010}.
Scalability is not only about computational efficiency, but also about lower communication overhead (e.g., how often a device needs to communicate with the remote cloud center), as well as reduced information needed (e.g., what type of information a device needs before making sensible decisions).

Faced with these major IoT challenges, innovations in network design and management are desired to enable efficient online operations, and seamless co-existence of humans with things \cite{chiang2016}. 
Consequently, it is imperative to develop new tools for IoT management that tap into diverse inference, signal processing, communications, and networking techniques, by drawing from fields such as machine learning, optimization, and applied statistics. The novel expertise gleaned from these research areas, coupled with the solid analytical approach, are the best credentials for succeeding in IoT research \cite{stankovic2014}.

From a network architecture perspective, to ensure the desired user experience and meet heterogeneous service requirements, IoT tasks nowadays are no longer only supported by the cloud data centers, but also through a promising new architecture termed \emph{edge computing}, or in a broader sense \emph{fog computing}. This architecture distributes computation, communication, and storage closer to the end IoT devices and users, along the cloud-to-things continuum \cite{chiang2016,barbarossa2018edge,barbarossa2014communicating,mach2017,wang2018twc,lyu2017}. 
This shift of computing paradigms is further promoted by the advanced communication techniques emerging with standards such as Narrowband-IoT (NB-IoT)~\cite{NB-IoT15,akpakwu2018survey}.

Given the huge volume of data in various IoT setups and the proliferation of learning and large-scale optimization advances, a pertinent direction is prompted by asking the following question: \emph{Can we learn from historical data to improve the quality of network management policies in IoT?} The rationale is that historical data contain statistics of the IoT environments \cite{vapnik2013}, and learning from them can mitigate the uncertainty of future management tasks. Further armed with online adaptation capability to reinforce the current policies, it is envisioned that \emph{learn-and-adapt} network management schemes can markedly improve IoT user experience in terms of low service delay, high system resilience, and adaptivity \cite{chen2017tcns,chen2016tsp}.
Toward this goal, the present overview paper will outline an offline-aided online approach with markedly improved performance, by leveraging statistical learning from historical samples.

Taking a step further, online learning, with online convex optimization (OCO) as a special case, is an emerging methodology for sequential decision making with {light-weight implementation} and well documented merits, especially when the environment (e.g., a sequence of convex costs) varies in an unknown and possibly adversarial manner \cite{zinkevich2003,cesa2006}.
Targeting a scalable solution in a prohibitively complex IoT environment, this paper will also overview a new OCO framework designed for IoT, which {further} incorporates various forms of feedback, physical constraints and performance metrics driven by IoT applications, relative to the standard settings \cite{zinkevich2003,cesa2006,hazan2016}.
Novel schemes tailored for this setting can lay a solid analytical foundation to delineate {the tradeoffs among algorithm scalability, performance guarantees,} and degree of (non-)stationarity present in the IoT environment \cite{chen2017tsp,chen2017iot}.

The rest of the paper is organized as follows. 
Section II deals with the heterogeneity in IoT demand and QoS, along with a unified formulation for dynamic IoT tasks.
Section III introduces methods for optimizing IoT performance under different level of non-stationarity in IoT dynamics.
Section IV summarizes scalable OCO-based schemes with different feedback options. Finally, concluding remarks and possible future research directions are highlighted in Section V.

\vspace{0.1cm}
\noindent\textbf{Notation}. Bold uppercase (lowercase) letters denote
matrices (column vectors), while $(\cdot)^{\top}$ stands for transposition, and $\|\mathbf{x}\|$ denotes the $\ell_2$-norm of a vector $\mathbf{x}$. The projection $[\mathbf{a}]^+:=\max\{\mathbf{a},\mathbf{0}\}$ are defined entrywise. 
The indicator function $\mathds{1}(A)$ takes value $1$ when the event $A$ happens, and $0$ otherwise. ${\cal O}(x)$ denotes big order of $x$, i.e., ${\cal O}(x)/x\rightarrow 1$ as $x\rightarrow 0$; $\tilde{\cal O}$ neglects the lower-order terms with a polynomial $\log x$ rate; and $\mathbf{o}(x)$ denotes small order of $x$, i.e., $\mathbf{o}(x)/x\rightarrow 0$ as $x\rightarrow 0$.

\section{Heterogeneity in IoT demand and QoS}
\label{sec:heter}
Heterogeneity is inherent in IoT, and it manifests itself across different aspects, from application requirements and constraints to sensing and communication technologies.
\subsection{Heterogeneous applications}
The range of IoT applications already spans several fields, and it is rapidly increasing. 
A few examples of applications are \cite{zanella2014internet,akpakwu2018survey,al2015internet}: (i) lifestyles (wearable gadgets, gaming, augmented/virtual reality, wellness); (ii) smart environments (homes, offices, cities); (iii) automotive (self-driving, traffic monitoring, intelligent transportation systems, vehicle-to-vehicle communications); (iv) industrial (full automation and control, structure monitoring, logistic); (v) environmental monitoring (pollution, global warming, waste management); (vi) healthcare (patient monitoring, body area networks, smart health, elderly care); and, (vii) security and surveillance. These applications are characterized by highly diverse requirements, in terms of data rate, latency, reliability, security, connectivity, mobility, etc. 
To illustrate the extreme variability of requirements, we note that virtual reality require latencies in the order of a few milliseconds and data rates in the order of $25$ Mbps, while automated driving or certain industrial control applications require latencies in the order of milliseconds and high packet transmission reliability (in the order of $99.999$ percent). Conversely, for  environmental monitoring applications such as waste management,  an update frequency of one packet/hour is sufficient, with a tolerable delay of $30$ minutes.

{
A few paradigms are useful to outline the challenges facing IoT, and the potential of our approaches to addressing them.

\vspace{0.1cm}
\noindent{\bf Automated driving.} The goal in this application is to enhance perception of an individual vehicle and thus improve safety. A common approach is to set up a {\it cooperative perception} system building on the information sharing between vehicles and roadside units (RSUs) \cite{Sakaguchi2018}. The scope is to widen the visibility of the individual vehicle to prevent that an object unseen by a single vehicle might cause an accident \cite{kim2015multivehicle,sakaguchi2017and}. The signals to be exchanged go from (low data rate) range measurements to (high data rate) high definition maps generated by sensors mounted on each vehicle. 
The communication channels between vehicles are highly dynamic and hard to predict, while the information available at each time slot can be outdated. Nevertheless, the communication among vehicles and RSUs should be performed in a reliable and timely manner to ensure that emergency operations can take place in the due time.

\vspace{0.1cm}
\noindent{\bf Intelligent transportation.} A typical task in this application is to navigate a set of electric vehicles to their destination, by collecting along the way data about traffic, state of the battery, and availability of parking slots. The objective here is to minimize fuel consumption and the time needed to reach a certain destination, while ensuring that all vehicles find a proper refueling station for their batteries along the way. The remaining time to destination is updated online, depending on the time-varying traffic state, which is generally unpredictable.} 

\subsection{Heterogeneous technologies}
The IoT ecosystem is composed of various components, whose functionality falls within the following categories \cite{al2015internet}: identification, sensing, communication, computation, and services. {\it Identification} is crucial to assign a clear identity to each object in the network. The role of {\it sensing} elements is to gather data from the real world. Typically, sensors are integrated with single board computers and TCP/IP functionalities to create IoT devices, such as Arduino or Raspberry PI, which are able to sense and send data to a decision entity. The role of \emph{communication} is to propagate information from the sensing elements to a decision entity, possibly distributed, and back to actuators. There is a plethora of very heterogeneous communication technologies that are in use in IoT. As a broad classification, we can list: (i) short-range technologies to support machine-to-machine communications, like Bluetooth, IEEE 802.15.4 and ZigBee; (ii) long-range networks, like LoRa supporting data rates of around $50$ kbps over ranges up to $15$ Km, or SigFox using ultra-narrowband technologies to support ultra-low power consumption and long ranges (up to 30-50 Km in rural areas, 3-10 Km in urban areas), at the expenses of limited data rates; (iii) Low-power Wi-Fi, also called IEEE 802.11ah  supporting data rates up to 347 Mbps; and, (iv) cellular networks. Current 4G cellular technologies, more specifically the $3$rd Generation Partnership Project Long-Term Evolution (3GPP LTE), represent the state-of-the art in mobile communications. However, LTE has been primarily designed for broadband communications, and thus not optimized for the machine-type communications (MTC) envisioned in IoT.

To partially overcome this discrepancy, 3GPP has introduced some modifications to the standards to enable the deployment of massive smart connected devices and services such as those in smart cities or smart grids: eMTC \cite{eMTC15} and NB-IoT \cite{NB-IoT15}. But a truly disruptive framework enabling an effective deployment of IoT is 5G communication networks, thanks to increased data
rate, reduced end-to-end latency, and improved coverage relative to 4G \cite{palattella2016internet}. The key features of 5G that
are particularly suitable for IoT are: (i) the integration of heterogeneous access technologies; (ii) virtualization of network functionalities; and, (iii) bringing cloud functionalities close to the end-user by introducing mobile edge computing (MEC). While earlier network generations have been designed as general purpose connectivity platforms, the vision underlying 5G is to create an ecosystem  for technical and business innovations involving vertical markets such as automotive, energy, agriculture, city management, healthcare, manufacturing, and transportation. Since these services have very different requirements and constraints, the key challenge of 5G is to design a single platform being able to serve different purposes in an efficient way.
 The solution to tackle such a challenging question is {\it network slicing}. 
 At the basis of network slicing, there is network function virtualization, which makes it possible to partition a single {\it physical} network into multiple {\it virtual} networks, each matched to its specific requirements and constraints. This enables operators to provide networks on an as-a-service basis, while meeting a wide range of use cases in parallel. Virtualization is going to play a key role also in IoT to cope with high heterogeneity of requirements as well as the capabilities of devices. But being able to meet the stringent latency requirements of IoT applications, virtualization needs to be coupled with a new architectural vision, enabled by MEC.

\begin{table*}[t]
\small
\begin{subequations}
\label{eq.allprob}
\begin{empheq}[box=\fbox]{align}
~~~\opt_{\{\mathbf{x}_t,\forall t\}}~~~~~~~~&\sum_{t=1}^Tf(\mathbf{x}_t;\mathbf{s}_t)\qquad\qquad~~~~~\//\texttt{IoT performance metrics}\label{eq.allproba}\\
\st~~~~~~&\sum_{t=1}^T\mathbf{g}(\mathbf{x}_t;\mathbf{s}_t)\leq \mathbf{0}\qquad\quad~~\//\texttt{IoT long-term requirements}\label{eq.allprobb}\\
&\,\mathbf{x}_t\in {\cal X}(\mathbf{s}_t),\,\forall t\qquad\qquad~~\//\texttt{IoT short-term requirements}\label{eq.allprobc}\\
&\,\texttt{Per slot $t$, IoT state dynamics}~\mathbf{s}_{t+1}=\mathbf{d}(\mathbf{s}_t,\mathbf{x}_t,\bm{\xi}_t)\label{eq.allprobd}\\
&\,\texttt{Per slot $t$, find $\mathbf{x}_t$ given information oracle ${\cal O}_t$}~~~\label{eq.allprobe}
\end{empheq}
\end{subequations}
\vspace{-0.4cm}
\end{table*}

\subsection{Embedding IoT in the edge cloud}\label{sec.mec}
Even within the sophisticated architecture of 5G networks, meeting the stringent latency constraints required in some IoT applications over a wide area network can be still challenging, if not impossible. 
To guarantee low latencies, a popular solution is to bring cloud functionalities close to the end users through mobile (or multi-access) edge computing \cite{hu2015mobile,liang2017mobile}. 

With MEC, computation and storage resources are brought at the edge of the network, represented by the network access points. 
In this way, delay-sensitive applications launched by a mobile device can be offloaded to the nearest mobile edge host (MEH), and the most popular contents can also be cached in MEHs to minimize downloading time \cite{chu2016allocating,chu2018}. 
Bringing computation and storage resources at the edge of the network makes it possible to guarantee low and stable delays. 
In practice, the applications launched by the user are executed by virtual machines running on nearby edge nodes, either cloudlets, exploiting a Wi-Fi connection \cite{cuervo2010maui}, \cite{kosta2012thinkair}, or MEHs, using cellular communication technologies \cite{barbarossa2014communicating}. 
The further extension of MEC is {\it fog computing}, where the edge of the network can include devices as well, thus creating a continuum of devices able to sense, communicate and compute \cite{bonomi2012fog,chiang2016}. 
A critical aspect in this scenario is mobility management \cite{sun2017emm}. 
To handle mobility while offering a seamless service continuity, it is necessary to migrate virtual machines quickly across MEH. 
This is a critical step, because instantiating a conventional virtual machine can take times well beyond the latencies required in some IoT applications. 

In MEC or fog computing settings, communication, computation and storage resources can be seen as three aspects of a single system. From a user-centric perspective, what actually matters is the time needed to launch an application and receive the result back. 
The overall delay depends on communication time, computation time and the distribution of contents across the network. 
This holistic vision calls for a {\it joint dynamic optimization of communication, computation, and caching resources} \cite{barbarossa2018edge}. 
An application where communication and computation resources are closely mingled is computation offloading. 
This is a fundamental mechanism to enable simple devices to run sophisticated applications or to allow battery-powered devices to run their applications remotely to save energy and thus prolong battery lifetime. 
Computation offloading has gained growing popularity recently. 
For single-user MEC, it has been studied in \cite{liu2016delay,kwak2015dream,lyu2017,mao2016dynamic}. 
The multi-user case was addressed in \cite{sardellitti2015joint,chen2016efficient}, and later extended to the dynamic case, using stochastic optimization in \cite{mao2017stochastic,stefania_icc2018}. See recent surveys \cite{barbarossa2018edge,barbarossa2014communicating,mao2017} and references therein.


\vspace{-0.2cm}

\subsection{Taming heterogeneity via a unified formulation}
{With various applications and technologies in mind, the goal of this section is to put forth a unified model for IoT tasks that will guide subsequent algorithmic development.} 
\vspace{0.05cm}

\noindent\textbf{Unifying models.}
Consider discrete time $t\in\mathbb{N}$, and a time horizon of $T$ slots. 
Per slot $t$, an IoT state variable $\mathbf{s}_t\in\mathbb{R}^p$ is defined, which characterizes all the critical parameters of the IoT environment.
Assuming certain amount of knowledge about the environment, the IoT operator will make a decision $\mathbf{x}_t\in\mathbb{R}^d$, aiming to optimize task-specific performance, subject to different types of constraints. 
The decision $\mathbf{x}_t$ can in turn \emph{drive} the next state $\mathbf{s}_{t+1}$. 
To model such decision making processes, we consider a generic problem \eqref{eq.allprob}. 
The model here is general. 
{The slot duration can vary from tens of microseconds in wireless networks, a few milliseconds in automated driving, tens of seconds in intelligent transportation}, to minutes or even hours in smart power networks; the state $\mathbf{s}_t$ can represent the channel gain in wireless networks, the congestion level in data networks as well as transportation networks, and the renewable generation, and energy prices in power networks; and the decision $\mathbf{x}_t$ can include the transmitted power in communication, the size of data workloads, the number of vehicles, or the amount of energy.
Regarding the objectives, constraints and state dynamics in \eqref{eq.allprob}, we will highlight their IoT relevance, especially of interest to communication and networking communities. 
\vspace{0.05cm}

\begin{table*}[t]
\small
\centering 
\caption{{An overview of heterogeneous IoT settings considered.}}\label{tab:overview} 
    \begin{tabular}{ c *{2}{|c}}
    \hline \hline
~~~~~~Section~~~~~~   & ~~~State dynamics $\mathbf{s}_t\rightarrow\mathbf{s}_{t+1}$~~~&Information oracle ${\cal O}_t$\\ \hline
Section \ref{sec.saopt1}    & i.i.d. or Markovian      & $f,\mathbf{g},{\cal X}$ and $\{\mathbf{s}_1\cdots,\mathbf{s}_t\}$     \\
Section \ref{sec.saopt2} & Partially controlled Markovian & $f,\mathbf{g},{\cal X}$ and $\{\mathbf{s}_1\cdots,\mathbf{s}_t\}$ \\
Section \ref{sec.saopt3}   & Controlled Markovian     & {$f_1(\mathbf{x}_1),\mathbf{g}_1(\mathbf{x}_1),\cdots,f_{t-1}(\mathbf{x}_{t-1}),\mathbf{g}_{t-1}(\mathbf{x}_{t-1})$ and $\{\mathbf{s}_1\cdots,\mathbf{s}_t\}$ } \\ \hline
Section \ref{sec:scale-1}    & Generally non-stationary  & $ f_1,\mathbf{g}_1,\cdots,f_{t-1},\mathbf{g}_{t-1},{\cal X}$  \\
Section \ref{sec:scale-2}   & Generally non-stationary   & $f_1(\mathbf{x}_1),\mathbf{g}_1,\cdots,f_{t-1}(\mathbf{x}_{t-1}),\mathbf{g}_{t-1},{\cal X}$    \\
Section \ref{sec:scale-3} & Generally non-stationary  & $f_1(\mathbf{x}_1),\mathbf{g}_1(\mathbf{x}_1),\cdots,f_{t-1}(\mathbf{x}_{t-1}),\mathbf{g}_{t-1}(\mathbf{x}_{t-1}),{\cal X}$    \\
 \hline \hline
    \end{tabular} 
    \vspace{-0.4cm}
\end{table*}

\noindent\textbf{Performance metrics.}
Given the state $\mathbf{s}_t$ and the decision $\mathbf{x}_t$, we consider the IoT performance as a generic time-invariant function $f(\mathbf{x}_t;\mathbf{s}_t)$ (use $f_t(\mathbf{x}_t)$ interchangeably) depending on the time-varying quantities $\mathbf{s}_t$ and $\mathbf{x}_t$. 
For MEC problem in Section \ref{sec.mec}, $f(\mathbf{x}_t;\mathbf{s}_t)$ often represents the power consumption aggregating over all devices, the aggregated delay, or the system throughput \cite{barbarossa2014communicating,wang2018twc,chen2017iot}. 
{
Another line of recent research studies a new performance metric in MEC --- age of information or age, which measures the timeliness of system status using the elapsed time since the most recently received packet was generated at its source \cite{kaul2012info}. Age of information is pertinent to mission-critical IoT applications \cite{costa2016tit,sun2017tit}.
Furthermore, for traffic assignment tasks in intelligent transportation, $f_t(\mathbf{x}_t)$ can capture the overall fuel consumption, and the travel time of vehicles on the road \cite{zhang2018pieee};} for demand response in smart grids, it is related to user utility and power balancing cost depending on the real-time energy prices \cite{lina2011,shi2017tsg,li2017tsg,wang2016jsac}; and for applications related to wireless communications, throughput or achievable rate also plays a critical role in the objective. 
\vspace{0.05cm}

\noindent\textbf{Short-term constraints.} 
The heterogeneous requirements in IoT are modeled via \emph{short-term} and \emph{long-term} constraints in \eqref{eq.allprob}. 
{The short-term constraints are imposed to regulate $\mathbf{x}_t$ in accordance to short-term requirements, which can be collected in a compact set ${\cal X}(\mathbf{s}_t)$ --- that is either continuous or discrete, and possibly depends on the IoT state $\mathbf{s}_t$. As an example, consider a MEC system composed of access points (APs), MEC servers, and mobile user equipment (UE). To meet the stringent latency requirement, the E2E latency of each UE should be less than $\bar{l}$, that is, 
\begin{equation}\label{Avg_Delay}
l_t^{\texttt{tx}}+l_t^{\rm bk}+l_t^{\texttt{exe}}+l_t^{\texttt{rx}}\leq \bar{l}
\end{equation}
where i) $l_t^{\texttt{tx}}$ is the time spent to send the program state and input (encoded with $b_t$ bits) from UE to AP, e.g., $l_t^{\texttt{tx}}=\displaystyle b_{t}/r_t$, 
with $r_t$ being the data rate (in bits/sec);  
ii) $l_t^{\rm bk}$ is the backhaul latency between AP and MEC server, which appears when the computations are performed in a server that is not co-located with the AP; iii) $l_t^{\texttt{exe}}$ is the server execution time defined as $l_t^{\texttt{exe}}=c_t/u_t$, 
where $c_t$ is the number of CPU cycles to be executed, and $u_t$ is the number of CPU cycles/second allocated by the MEC server to UE; and, iv) $l_t^{\texttt{rx}}$ is the time for the MEC server to send back the result to UE.} 
With $\mathbf{x}_t:=\{u_t, r_t\}$ and $\mathbf{s}_t:=\{l_t^{\rm bk}, l_t^{\texttt{rx}},c_t, b_t\}$ thus \eqref{Avg_Delay} included in ${\cal X}(\mathbf{s}_t)$, selecting $\mathbf{x}_t\in {\cal X}(\mathbf{s}_t)$ guarantees the E2E latency requirement in MEC. Short-term constraints also arise due to the physical limits of transmission lines and generators in power networks \cite{giannakis2013}, transceivers in wireless communication \cite{wang2007}, as well as vehicles in transportation networks \cite{zhang2018pieee}. 
\vspace{0.05cm}

\noindent\textbf{Long-term constraints.}
In some IoT applications, the short-term constraints cannot accurately characterize the demand and requirements. 
{For the latency requirement in MEC, the short-term constraint \eqref{Avg_Delay} makes implicit assumptions that i) no new task is generated before the old tasks are completed; and, ii) each single task is carried out within an established time frame. 
These assumptions may be restrictive in some cases.  
Consider also a vehicle in the intelligent transportation application that must arrive at its destination within a certain interval. To guarantee on-time arrival, its long-term average speed instead of the instantaneous speed needs to be lower bounded.} 
The long-term constraints are thus well-motivated to allow flexible adaptation of $\mathbf{x}_t$ to temporal variations of service requirements. 
Given the state $\mathbf{s}_t$ and the decision $\mathbf{x}_t$, they are modeled as 
a set of penalty functions $\mathbf{g}(\mathbf{x}_t;\mathbf{s}_t):=[g^1(\mathbf{x}_t;\mathbf{s}_t),\cdots,g^N(\mathbf{x}_t;\mathbf{s}_t)]^{\top}$ in \eqref{eq.allprobb}. 
Ideally, we want the accumulated penalty over the entire horizon below a certain threshold. For convenience, we let the threshold to be $\mathbf{0}$ in \eqref{eq.allprobb}, which is without loss of generality subject to a constant shift.
Long-term constraints also appear in wireless networks where often the average transmit power and link capacity are confined \cite{wang2007}. 
The challenge in dealing with long-term constraints is that the future states $\mathbf{s}_{t+1},\cdots,\mathbf{s}_T$ are not known at slot $t$, which calls for adaptive optimization.
\vspace{0.05cm}

\noindent\textbf{State dynamics.}
One of the key challenges in IoT is its unpredictable dynamics. 
In \eqref{eq.allprob}, IoT dynamics are encoded by a state transition function $\mathbf{d}$ which generates the next state $\mathbf{s}_{t+1}=\mathbf{d}(\mathbf{s}_t,\mathbf{x}_t,\bm{\xi}_t)$ given $\mathbf{s}_t$ and $\mathbf{x}_t$ as well as an exogenous variable $\bm{\xi}_t$. In most cases, the exogenous variable $\bm{\xi}_t$ can be a random disturbance.
For wireless communication applications where the state $\mathbf{s}_t$ represents the fading channel state, then $\mathbf{s}_{t+1}$ often does not depend on $\mathbf{s}_t$ and $\mathbf{x}_t$; that is, $\mathbf{s}_{t+1}=\mathbf{d}(\bm{\xi}_t):=\bar{\mathbf{s}}+\bm{\xi}_t$, where $\bar{\mathbf{s}}$ is the mean channel state, and $\bm{\xi}_1,\cdots,\bm{\xi}_T$ are independent, identically distributed (i.i.d.) zero-mean random variables; see e.g., \cite{wang2007,gatsis2010}. 
Markovian dynamics are also common in modeling energy prices, renewable generation processes \cite{gonzalez2005}, in which case $\mathbf{s}_{t+1}=\mathbf{d}(\mathbf{s}_t,\bm{\xi}_t)$ depends on the current state $\mathbf{s}_t$ and an i.i.d. noise $\bm{\xi}_t$ but not $\mathbf{x}_t$. 
{We refer to both $\mathbf{s}_{t+1}=\mathbf{d}(\bm{\xi}_t)$ and $\mathbf{s}_{t+1}=\mathbf{d}(\mathbf{s}_t,\bm{\xi}_t)$, as \emph{non-interactive} dynamics.} 
The decision $\mathbf{x}_t$ can also play an important role in state transitions.
Taking MEC as an example, a queueing model is usually incorporated to keep trace of the relevant quantities, such as the amount of remaining tasks that need to be offloaded or processed. 
{With $\{b_t, c_t, r_t, u_t\}$ defined below \eqref{Avg_Delay}, we consider a transmission queue $q_t^{\rm tx}$ that quantifies the number of bits to be transmitted at slot $t$ from UE, and a computation queue  $q_t^{\rm exe}$ that quantifies the amount of computation that needs to be completed for UE. 
If $\Delta t$ denotes the slot duration, the transmission queue evolves as
\begin{equation}
q_{t+1}^{\rm tx} = \max\left[q_{t}^{\rm tx}-r_t \Delta t, 0\right] + b_{t}
\end{equation}
and the computation queue evolves as follows $q_{t+1}^{\rm exe} = \max\left[q_{t}^{\rm exe}-u_t\Delta t, 0\right] + c_{t}$.
In this case, the IoT state is $\mathbf{s}_t:=\{q_t^{\rm tx},q_t^{\rm exe}\}$, the decision is $\mathbf{x}_t:=\{u_t, r_t\}$, and the exogenous variable is $\bm{\xi}_t:=\{b_t,c_t\}$. It then follows that $\mathbf{s}_{t+1}=\mathbf{d}(\mathbf{s}_t,\mathbf{x}_t,\bm{\xi}_t)$ --- what we term \emph{interactive} dynamics, or more precisely, \emph{controlled} Markovian dynamics if $\bm{\xi}_t$ is i.i.d. 
If the communication and computation resources are sufficient, an ideal policy should guarantee the queue stability \cite{tassiulas1992,neely2010}.} 
According to Little's law \cite{ross2014introduction}, the average execution delay experienced by each UE is proportional to the average queue lengths. 
Hence, a meaningful problem can be minimizing the average power, subject to the average delay constraints, which will be discussed in Section \ref{sec.saopt2}.
State variables of this type also include the location of {a vehicle in the intelligent transportation} or an unmanned aerial vehicle (UAV) that depends on their previous location and the current movement \cite{gregoire2015,wu2018,xu2018}, and the energy level of a battery that depends on their instantaneous (dis)charging amounts. 
More complex dynamics are also possible in IoT due to e.g., strategic human interactions and malicious attack \cite{zou2016}. 
In those cases, $\bm{\xi}_t$ can be a function of all the states $\mathbf{s}_1,\cdots,\mathbf{s}_t$, or even completely arbitrary. 
\vspace{0.05cm}

\noindent\textbf{Accessible information.}
While various objectives, constraints and state dynamics have been adopted to model heterogeneous problems in IoT, the level of accessible information directly affects how to solve the resultant problem given \emph{limited} communication and computation resources --- the epicenter of scalability barriers in IoT. 
Let the information oracle ${\cal O}_t$ collect all the information available to the IoT operator before making decision $\mathbf{x}_t$. 
For cases where the objectives and the constraints are easy-to-measure formulas (e.g., aggregated power, throughput, distance), we consider ${\cal O}_t:=\{f,\mathbf{g},{\cal X},\mathbf{s}_1\cdots,\mathbf{s}_t\}$ that includes the explicit form of functions $\{f,\mathbf{g}\}$, set ${\cal X}$ and one-slot-ahead prediction $\mathbf{s}_t$. 
In some IoT settings however, i) the objective capturing user-centric quantities, e.g., service latency or reliability, security risk, and customer ratings, is hard to model; ii) the objective involving fast-varying quantities is hard to predict, e.g., the millimeter wave links in 5G are prone to blocking events, thus hard to predict.; and, iii) even if modeling and predicting are possible in theory, the low-power smart devices may not afford the complexity of running statistical learning tools ``on-the-fly.'' 
In such cases, we consider a fully causal information oracle ${\cal O}_t:=\{f_1(\mathbf{x}_1),\mathbf{g}_1(\mathbf{x}_1),\cdots,f_{t-1}(\mathbf{x}_{t-1}),\mathbf{g}_{t-1}(\mathbf{x}_{t-1})\}$ that includes only the observed objective function \emph{values} and constraint \emph{penalties} at previous slots. 
IoT scenarios between these two extreme cases will also be discussed. 

{In Table \ref{tab:overview}, we summarize the heterogeneous settings that one may encounter in IoT. 
Targeting these settings, a set of suitable solvers will be discussed in the subsequent sections.   
While the methodologies presented in this paper mainly focus on stochastic optimization and online learning, approaches based on other methodologies such as game theory and robust optimization can be also applied to solve similar problems. 
}



\section{Adaptivity to dynamic IoT environments}\label{sec:adapt}
\begin{figure}
\centering
\includegraphics[width=0.46\textwidth]{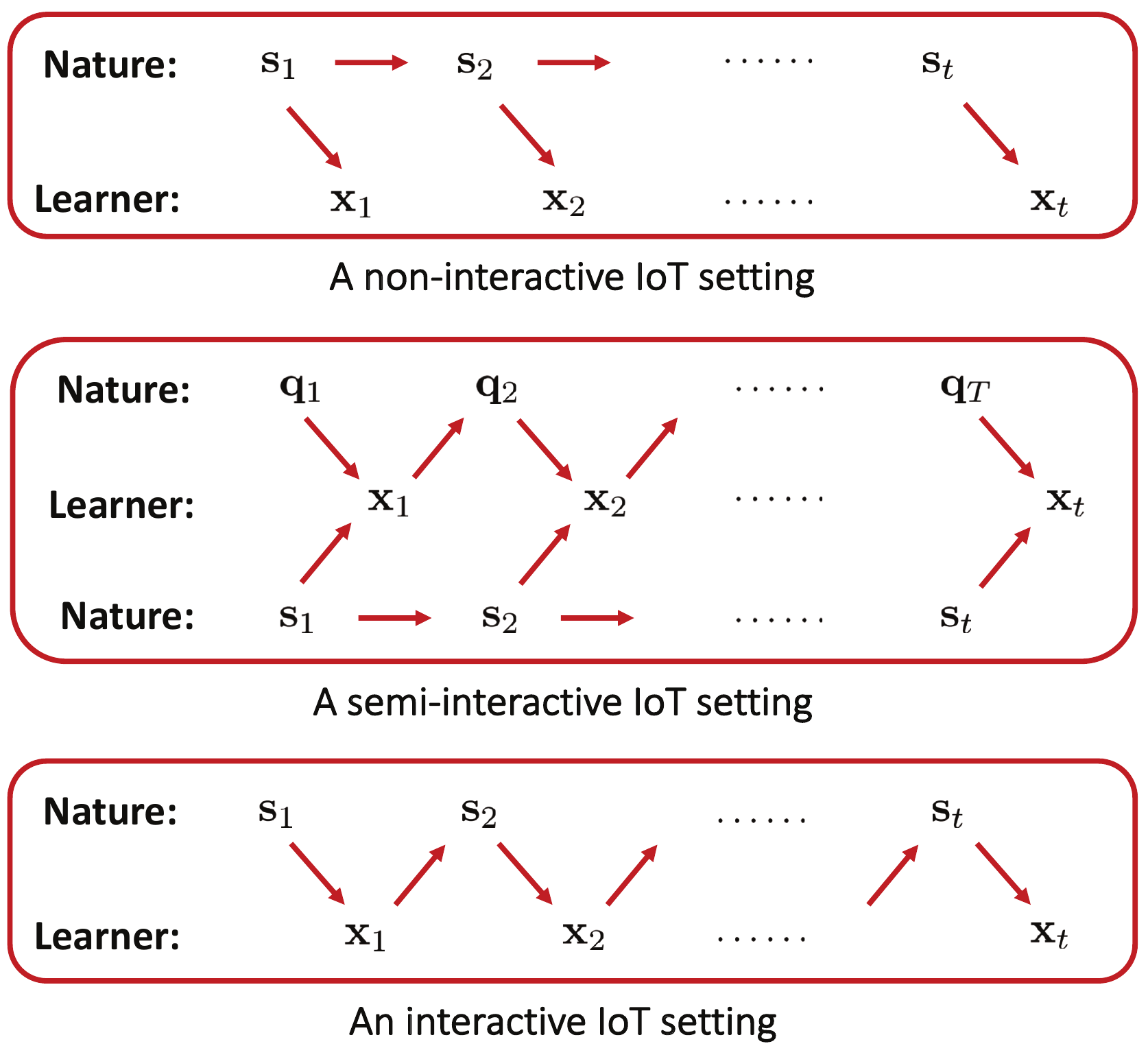}
\caption{\small {Three levels of interaction between IoT operator (learner) and nature corresponding to the three assumptions in Table \ref{tab:overview}.}} \label{fig:saopt-comp}
\vspace{-0.1cm}
\end{figure}

This section introduces methods for optimizing IoT performance under the (asymptotically) 
stationary assumption on IoT dynamics relative to control decisions in the fast timescale.
Corresponding to different types of state dynamics in Table \ref{tab:overview}, we outline three classes of management schemes; see Fig. \ref{fig:saopt-comp}.

\vspace{-0.1cm}
\subsection{Leveraging statistical learning for IoT management}\label{sec.saopt1}
As the generic problem \eqref{eq.allprob}, consider the IoT operator makes a per-slot decision $\mathbf{x}_t$, subject to the short-term constraints that are collected in a compact set ${\cal X}(\mathbf{s}_t)$ parameterized by the IoT state $\mathbf{s}_t\in{\cal S}$, as well as the long-term constraints that are expressed as a time-varying penalty function
$\mathbf{g}(\mathbf{x}_t;\mathbf{s}_t)\in\mathbb{R}^N$. 
With the IoT cost $f(\mathbf{x}_t;\mathbf{s}_t)$, we wish to find a sequence of decisions $\{\mathbf{x}_t\}$ that minimize the expected limiting-average cost subject to the long-term and short-term constraints, i.e.,\begin{subequations}
\label{eq.prob-saopt0}
\begin{align}
f^*\!:=\!\!\!\mini_{\{\mathbf{x}_t\in {\cal X}(\mathbf{s}_t),\,\forall t\}}~&\lim_{T\rightarrow\infty}\frac{1}{T}\sum_{t=1}^T\mathbb{E}\left[f(\mathbf{x}_t;\mathbf{s}_t)\right]  \label{eq.prob-saopt0a}~~\\
\text{subject to}~~  &\lim_{T\rightarrow\infty}\frac{1}{T}\sum_{t=1}^T\mathbb{E}\left[\mathbf{g}(\mathbf{x}_t;\mathbf{s}_t)\right]\leq \mathbf{0}\label{eq.prob-saopt0b}
\end{align}
\end{subequations}
where $\mathbb{E}$ is taken over the random state $\mathbf{s}_t$, and possible randomness we may opt to introduce in the decision $\mathbf{x}_t$. 
Comparing with \eqref{eq.allprob}, the infinite time horizon and the limiting average cost are used in \eqref{eq.prob-saopt0} and throughout this section for mathematical simplicity. 
Indeed, assuming $\mathbf{s}_t$ is i.i.d. or generally stationary, the \emph{dynamic} problem \eqref{eq.prob-saopt0} shares the same optimal objective value as the following \emph{static} problem \cite{neely2010}
\begin{subequations}
\label{eq.prob-saopt}
\begin{empheq}[box=\fbox]{align}
f^*\!:=\!\!\!\mini_{\{\bm{\pi}(\mathbf{s}_t)\in {\cal X}(\mathbf{s}_t),\,\mathbf{s}_t\}}~&\mathbb{E}\left[f(\bm{\pi}(\mathbf{s}_t);\mathbf{s}_t)\right]  \label{eq.proba}~~\\
\text{subject to}~~  &\mathbb{E}\left[\mathbf{g}(\bm{\pi}(\mathbf{s}_t);\mathbf{s}_t)\right]\leq \mathbf{0}.\label{eq.probb}
\end{empheq}
\end{subequations}
To this end, our goal is to determine a possibly randomized policy $\bm{\pi}$ that given an IoT state $\mathbf{s}_t$, generates $\mathbf{x}_t=\bm{\pi}(\mathbf{s}_t)$ so as to minimize the average cost subject to both long- and short-term constraints in \eqref{eq.prob-saopt}.
The infinite-dimension \emph{functional} optimization problem \eqref{eq.prob-saopt} is more tractable in its dual form, which entails a finite number of variables \cite{chen2016tsp,Ale10}.
With $\bm{\lambda}\in \mathbb{R}_+^N$ denoting the multipliers, the Lagrangian of \eqref{eq.prob-saopt} is ${\cal L}(\bm{\pi},\bm{\lambda}) \!:=\mathbb{E}\big[{\cal L}(\bm{\pi}(\mathbf{s}_t),\bm{\lambda};\mathbf{s}_t)\big]$
where the instantaneous (per state) Lagrangian is ${\cal L}(\bm{\pi}(\mathbf{s}_t),\bm{\lambda};\mathbf{s}_t):=f(\bm{\pi}(\mathbf{s}_t);\mathbf{s}_t)+\bm{\lambda}^{\top}\mathbf{g}(\bm{\pi}(\mathbf{s}_t);\mathbf{s}_t)$.
Correspondingly, the dual problem of \eqref{eq.prob-saopt} is
\begin{align}\label{eq.dual-saopt}
 \maxi_{\bm{\lambda}\geq \mathbf{0}} ~~~ {\cal D}(\bm{\lambda}):=\mathbb{E}\left[{\cal
 D}(\bm{\lambda};\mathbf{s}_t)\right]
\end{align}
where ${\cal D}(\bm{\lambda};\mathbf{s}_t):=\!\min_{\mathbf{x} \in {\cal X}(\mathbf{s}_t)}{\cal L}(\mathbf{x},\bm{\lambda};\mathbf{s}_t)$.
With the optimal $\bm{\lambda}^*$ obtained for the \emph{dual problem} (\ref{eq.dual-saopt}), the optimal policy for the problem \eqref{eq.prob-saopt} could be retrieved as
\begin{equation}\label{eq.saopt-opt}
\bm{\pi}^*(\mathbf{s}_t):=\arg\min_{\mathbf{x} \in {\cal X}(\mathbf{s}_t)}{\cal L}(\mathbf{x},\bm{\lambda}^*;\mathbf{s}_t).	
\end{equation}

The \textit{ensemble} problem \eqref{eq.dual-saopt} is difficult to solve since the probability density distribution of $\mathbf{s}_t$ is usually unknown.
To find the optimal multipliers $\bm{\lambda}^*$ in an efficient manner, existing methods mainly rely on the stochastic subgradient-based (SGD) methods \cite{wang2007,gatsis2010,neely2010}.
However, SGD is known to suffer from slow convergence, which implies that the IoT network needs to implement sufficient many suboptimal decisions generated during the transient stage of SGD.

From a different viewpoint, given the huge volume of historical data generated by IoT networks, \eqref{eq.dual-saopt} was first formulated in \cite{chen2016tsp} as a statistical
learning task involving both {\em offline} training and {\em online} operational phases.
The rationale is that historical data contain statistics of the IoT states, and learning from them can aid coping with the uncertainty of future management tasks, leading to reduced transient time of adaptive algorithms.

Specifically, with a training set of $N_0$ historical IoT state samples $\hat{\cal S}_0:=\{\mathbf{s}_n,1\leq n\leq N_0\}$ available offline, \eqref{eq.dual-saopt} can be recast in an empirical form via sample averaging as
\begin{align}\label{eq.dual-saopt2}
\!\!\!\max_{\bm{\lambda}\geq \mathbf{0}} \, \hat{{\cal D}}_{\hat{\cal S}_0}(\bm{\lambda}),~{\rm with}~\hat{{\cal D}}_{\hat{\cal S}_0}(\bm{\lambda})\!:=\!\frac{1}{N_0}\sum_{n=1}^{N_0}\hat{{\cal D}}_n(\bm{\lambda})-\frac{\epsilon}{2}\|\bm{\lambda}\|^2\!
\end{align}
where $\hat{{\cal D}}_n(\bm{\lambda}):={\cal D}(\bm{\lambda};\mathbf{s}_n)$, and $\epsilon>0$ is a regularization constant typically used in statistical learning to boost generalization capability \cite{vapnik2013}.
{Note that while an $\ell_2$-norm regularizer is adopted in \eqref{eq.dual-saopt2}, other forms of regularization (e.g., $\ell_1$ and total-variation norm) are also possible depending on a-priori knowledge.}
Note that here $t$ has been replaced by $n$ to differentiate historical data from data in online phases.

Viewing \eqref{eq.dual-saopt2} as a (negated) empirical risk minimization (ERM) task,
we can resort to the state-of-the-art optimization methods for ERM, e.g.,
SAGA \cite{defazio2014}, that enjoys fast convergence and low complexity.
Using SAGA, per iteration $k$, we evaluate a single summand of the empirical gradient, i.e., $\nabla\hat{{\cal D}}_{\nu(k)}(\bm{\lambda}_k)$ at the iterate $\bm{\lambda}_k$, with sample index $\nu(k)\!\in\!\{1,\ldots,N_0\}$ selected \textit{uniformly at random}. Thus, the computational complexity of SAGA is that of a SGD iteration for \eqref{eq.dual-saopt2}.
Furthermore, SAGA stores a collection of the outdated gradients $\{\nabla_{\rm old}\hat{{\cal D}}_n\}$ for all samples, where $\nabla_{\rm old}\hat{{\cal D}}_n$ was evaluated by $\bm{\lambda}_{k[n]}$ --- the most recent iteration $k[n]$ that $\mathbf{s}_n$ was drawn; i.e., $k[n]\!:=\!\sup\{k'\!:\!\nu(k')\!=\!n, k'\!<\! k\}$. SAGA combines the fresh gradient with the stored ones as
\begin{equation}\label{eq.saga}
\bm{\lambda}_{k+1}\!=\!\Big[\bm{\lambda}_k+\alpha \Big(\nabla\hat{{\cal D}}_{\nu(k)}(\bm{\lambda}_k)\!-\!\nabla_{\rm old}\hat{{\cal D}}_{\nu(k)}+\nabla_{\rm old}\hat{\cal D}_{\hat{\cal S}_0}\Big)\Big]^{+}\!\!
 \end{equation}
where $\alpha$ is the pre-defined stepsize, and the stored gradients are $\nabla_{\rm old}\hat{\cal D}_{\hat{\cal S}_0}:=({1}/{N})\textstyle\sum_{n=1}^N\nabla_{\rm old}\hat{{\cal D}}_n-\epsilon \bm{\lambda}_k$.

\begin{figure}[t]
\centering
\vspace{-0.1cm}
\includegraphics[width=0.45\textwidth]{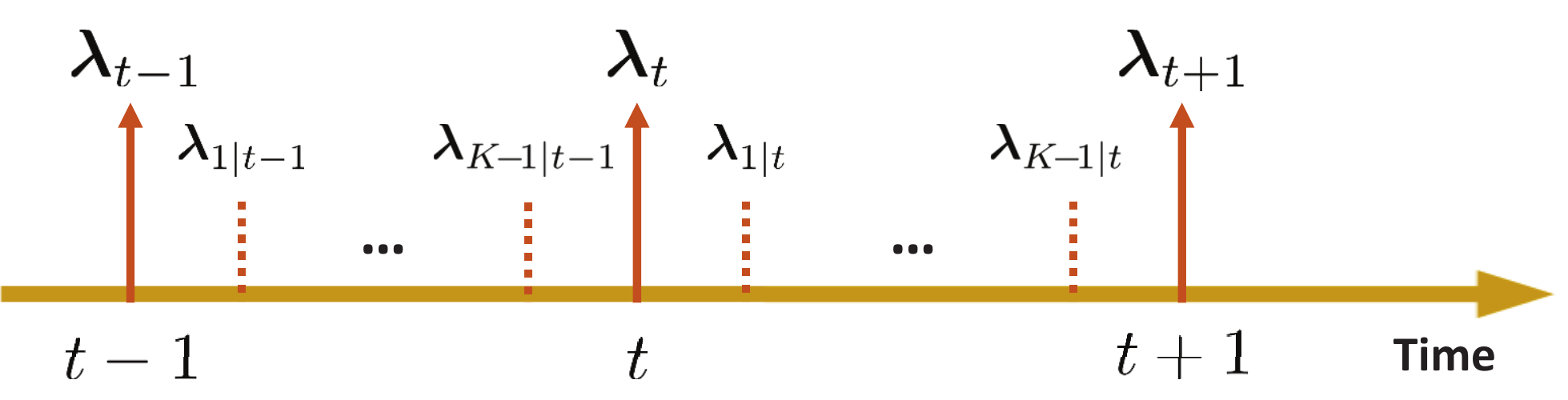}
\caption{\small Timescale splitting for offline-aided online SAGA operations. Iterates $\{\bm{\lambda}_t\}$ generate actual IoT decision, while $\bm{\lambda}_{1|t},\ldots,\bm{\lambda}_{K|t}$ are $K$ \emph{virtual} iterates updated via \eqref{eq.saga} at slot $t$; and $\bm{\lambda}_{t+1}:=\bm{\lambda}_{K|t}$.} \label{fig:saga-IoT}
\vspace{-0.1cm}
\end{figure}

The merits of SAGA lie in the fact that its gradient estimator in \eqref{eq.saga} is still unbiased as that with SGD.
In addition to the unbiasedness however, SAGA's gradient estimator attains considerably lower variance than SGD thanks to the contribution of the stored previous gradients, which is now termed the \emph{variance reduction} technique prevalent in large scale machine learning tasks.
Needless to mention the encouraging empirical results, the SAGA in \eqref{eq.saga} is provably convergent to the optimum of \eqref{eq.dual-saopt2} with the linear convergence rate \cite{chen2016tsp,defazio2014}
\begin{equation}\label{eq.saga-rate}
\mathbb{E}_{\nu}\left[\hat{\cal D}_{\hat{\cal S}_0}^*-\hat{\cal D}_{\hat{\cal S}_0}(\bm{\lambda}_k)\right]={\cal O}\big(\rho^k\big)
\end{equation}
where $\hat{\cal D}_{\hat{\cal S}_0}^*$ is the optimal objective of \eqref{eq.dual-saopt2}, and $\rho\in(0,1)$ is the linear rate depending on the objective function of \eqref{eq.dual-saopt2}.

Hence, in the offline phase, we run $KN_0$ SAGA iterations \eqref{eq.saga} on set $\hat{\cal S}_0$ - on average $K$ iterations per sample.
In the online phase, initialized with the offline output, SAGA (we term online SAGA) keeps acquiring data $\mathbf{s}_t$ with a growing training set $\hat{\cal S}_t:=\hat{\cal S}_{t-1}\cup\mathbf{s}_t$.
At slot $t$, online SAGA is initialized with the last iterate of slot $t-1$, and updates $\bm{\lambda}_t$ by running $K$ iterations \eqref{eq.saga}; see Fig. \ref{fig:saga-IoT}. The IoT decision is generated using the current $\bm{\lambda}_t$ by $\bm{\pi}_t(\mathbf{s}_t)=\arg\min_{\mathbf{x}\in {\cal X}(\mathbf{s}_t)}{\cal L}(\mathbf{x},\bm{\lambda}_t;\mathbf{s}_t)$.
This is the key idea of \emph{offline-aided online} IoT operations.


The offline-aided online scheme is \emph{not} simply heuristic.
In fact, the learning performance can be rigorously quantified via several concentration results in the learning theory \cite{vapnik2013}, which uniformly bound the discrepancy between the empirical loss \eqref{eq.dual-saopt2} and the loss \eqref{eq.dual-saopt} with high probability (whp), i.e.,
	\begin{equation}
		\sup_{\bm{\lambda}\geq \mathbf{0}}|{\cal D}(\bm{\lambda})-\hat{{\cal D}}_{\hat{\cal S}_t}(\bm{\lambda})|\leq {\cal H}_s(N_t),~{\rm whp}
	\end{equation}
where ${\cal H}_s(N_t)$ bounds the statistical error induced by the finite size $N_t$ of the training set $\hat{\cal S}_t$.
Under proper (so-termed mixing) conditions, the law of large numbers guarantees that ${\cal H}_s(N_t)$ is generally in the order of ${\cal O}(\sqrt{1/N_t})$ \cite[Section 3.4]{vapnik2013}. On the other hand, let ${\cal H}_o(KN_t)$ upper bound the optimization error of solving \eqref{eq.dual-saopt2} with $\hat{\cal S}_t$ due to running on average only finite ($K$) iterations per sample; i.e., $\hat{\cal D}_{\hat{\cal S}_t}^*\!\!-\!\hat{\cal D}_{\hat{\cal S}_t}(\bm{\lambda}_t)\leq {\cal H}_o(KN_t)$.

Online SAGA aims at a ``sweet-spot'' between affordable complexity (controlled by $K$) and desirable \textit{overall learning error}, which accounts for both the optimization and statistical errors ${\cal H}_s(N_t)+{\cal H}_o(KN_t)$.
Specifically, if we select $N_0\geq 3\kappa/4$ with $\kappa$ denoting the condition number of \eqref{eq.dual-saopt2}, and $K\geq 6$, the optimization error is bounded by ${\cal H}_o(KN_t)\leq {\cal H}_s(N_t)$ \cite{chen2016tsp}. In fact, even with $K=1$, online SAGA can still guarantee that ${\cal H}_o(KN_t)={\cal O}\left({\cal H}_s(N_t)\right)$.
With the link between the optimal policy and the optimal multiplier \eqref{eq.saopt-opt} in mind, the key message here is that with sufficient historical samples, online SAGA only requires running a small number of iterations per slot to bring the optimization error close to the statistical accuracy provided by the current training set.
Recent works along this line also include \cite{eisen2018,huang2015rlc} that focused on algorithms for \emph{piecewise stationary} environments. 
Learning more complex policies for non-interactive settings has been also studied by leveraging deep neural networks \cite{sun2017}. 
{Possible future research along this line also includes developing algorithms under the assumption of stationarity in high-order moments, which is also pertinent in practice. Algorithms tailored for fully nonstationary settings will be presented in Section \ref{sec:scale}.}

\vspace{-0.1cm}
\subsection{Learn-and-adapt approaches in semi-interactive settings}\label{sec.saopt2}
The IoT environment in Section \ref{sec.saopt1} is non-interactive, meaning that the dynamic of $\mathbf{s}_{t+1}$ in \eqref{eq.prob-saopt} does not change according to $\mathbf{x}_t$.
The IoT states can be also driven by decisions, which include the job queue length in a data center \cite{chen2016}, the lane length in a transportation network \cite{gregoire2015}, as well as the battery level in a smart grid \cite{shi2017tsg,wang2018tsg}. 
This section considers the case where such IoT states appear in the constraints, but not in the objectives, which we call the \emph{semi-interactive settings}.

Consider an IoT network represented as a directed graph ${\cal G}=({\cal N},\,{\cal E})$ with nodes ${\cal N}:=\{1,\ldots,N\}$ and edges ${\cal E}:=\{1,\ldots,E\}$.
The node-incidence matrix is formed with $(n,e)$ entry $\mathbf{A}_{(n,e)}\!=\!1(-1)$ if link $e$ enters (leaves) node $n$, and $\mathbf{A}_{(n,e)}=0$, otherwise.
With $\mathbf{c}_t\in \mathbb{R}_+^{N}$ collecting the \emph{exogenous} resources of all nodes per slot $t$, $\mathbf{x}_t\in \mathbb{R}^{E}$ for the \emph{endogenous} resources across edges, the aggregate resource is
$\mathbf{A}\mathbf{x}_t+\mathbf{c}_t$.
Connecting with \eqref{eq.prob-saopt}, $\mathbf{c}_t$ is included in the IoT state $\mathbf{s}_t$, and the constraint becomes $\mathbf{g}(\mathbf{x}_t;\mathbf{s}_t)=\mathbf{A}\mathbf{x}_t+\mathbf{c}_t$.
With $\mathbf{q}_{t}$ collecting all \emph{buffered} resources at slot $t$, we wish to solve \eqref{eq.prob-saopt0} with the additional state dynamics and the long-term constraints as
\begin{subequations}\label{eq.prob-saopt2}
\begin{align}
&\mathbf{q}_{t+1}=\left[\mathbf{q}_t+\mathbf{A}\mathbf{x}_t+\mathbf{c}_t\right]^{+}\!\!,\,\forall t\label{eq.prob-saopt2b}\\
~&\lim_{T\rightarrow \infty} ({1}/{T}) {\textstyle\sum_{t=1}^{T}} \mathbb{E}\left[\|\mathbf{q}_t\|\right]< \infty. \label{eq.prob-saopt2c}
\end{align}
\end{subequations}
{Due to the extra constraints in \eqref{eq.prob-saopt2}, the optimal objective of this new problem is at least $f^*$ in \eqref{eq.prob-saopt0}.}
Furthermore, the dynamic of the interactive state $\mathbf{q}_t$ (a.k.a. queues) in \eqref{eq.prob-saopt2b} also accounts for the transient performance of an adaptive algorithm.
To see this, suppose that under $\bm{\pi}^*$, it holds that $\mathbf{A}\bm{\pi}^*(\mathbf{s}_t)+\mathbf{c}_t=\mathbf{0},\,\forall \mathbf{s}_t$; and consider the convergence path of policy $\bm{\pi}_t$ induced by $\bm{\lambda}_t$ as $\bm{\pi}_1\rightarrow\bm{\pi}_2\rightarrow\bm{\pi}_3=\ldots=\bm{\pi}_T=\bm{\pi}^*$, along with $\mathbf{A}\bm{\pi}_1(\mathbf{s}_1)+\mathbf{c}_1=\mathbf{10}$ and $\mathbf{A}\bm{\pi}_2(\mathbf{s}_2)+\mathbf{c}_2=\mathbf{5}$. In this case, if $\mathbf{q}_1=\mathbf{0}$, then we have $\mathbf{q}_2=\mathbf{10}$ and $\mathbf{q}_3=\ldots=\mathbf{q}_T=\mathbf{15}$.

The simple example entails two variable insights: i) constraint violations incurred by the sub-optimal decisions during the transient stage (e.g., $\bm{\pi}_1, \bm{\pi}_2$) accumulate via $\mathbf{q}_t$; and ii) once accumulated in the transient stage, $\mathbf{q}_t$ will not decrease in the steady state (e.g., $\bm{\pi}_t,\,t\geq 3$). This explains the suboptimal performance tradeoff of SGD for \eqref{eq.prob-saopt0} with \eqref{eq.prob-saopt2}; see also \cite{chen2017tcns}.

\begin{figure}[t]
\centering
\vspace{-0.2cm}
\includegraphics[width=0.48\textwidth]{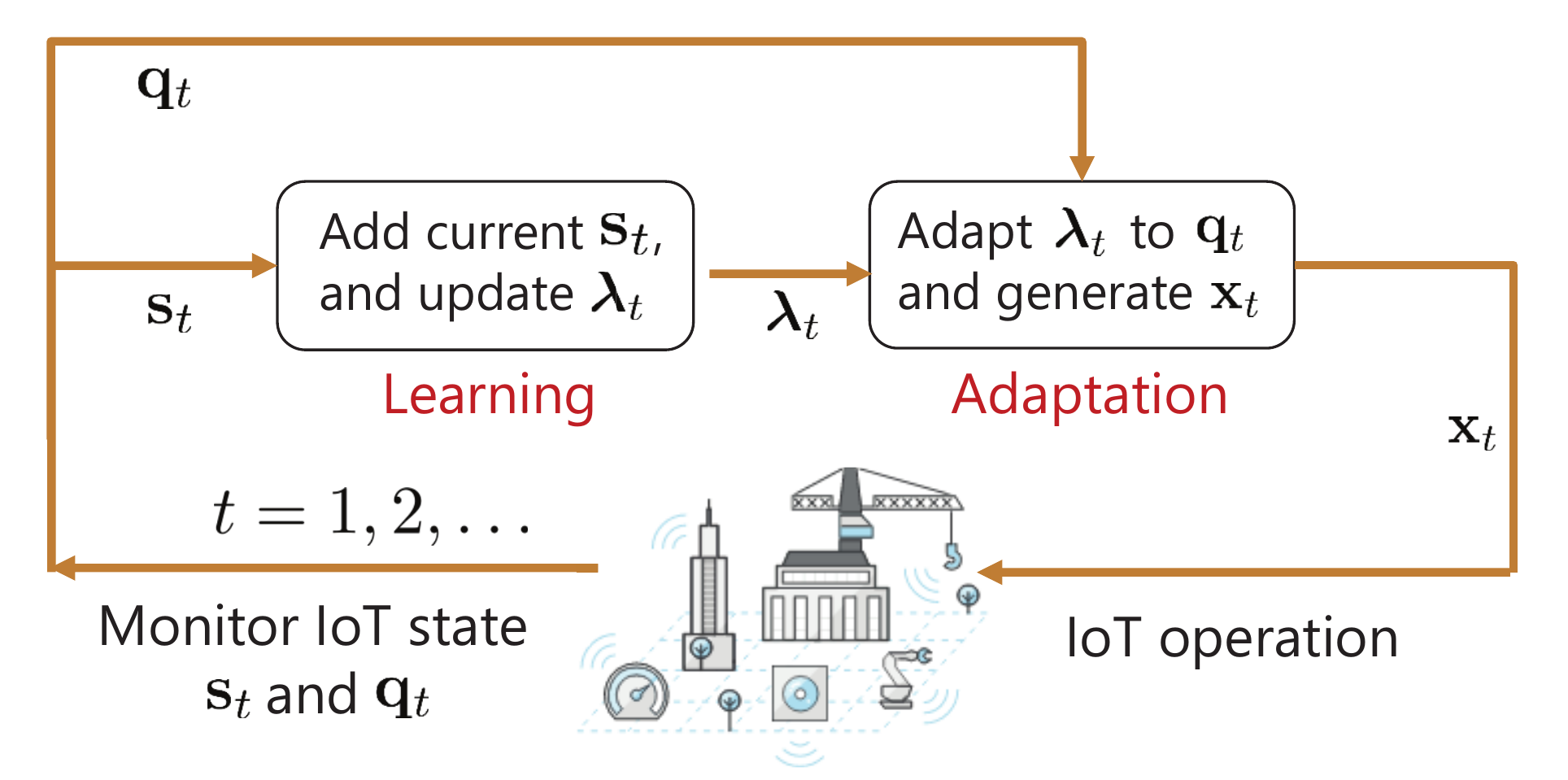}
\vspace{-0.2cm}
\caption{\small A learn-and-adapt diagram for IoT management.} \label{fig:LA-IoT}
\vspace{-0.2cm}
\end{figure}

To better control the interactive state $\mathbf{q}_t$, it suffices to:
\vspace{-0.05cm}

\noindent S1) reduce the transient time of the adaptive algorithm; and,
\vspace{-0.05cm}

\noindent S2) diminish $\mathbf{q}_t$ accumulated during the transient stage.

Following this two guidelines, we adopt a procedure in the online phase that we term \textit{online learning-and-adaptation} (named use LA-SAGA henceforth); see a diagram in Fig. \ref{fig:LA-IoT}.
Regarding S1), LA-SAGA performs the offline-aided online learning as that in Fig. \ref{fig:saga-IoT}, by growing the training set $\hat{\cal S}_t$ based on which it better learns $\bm{\lambda}_t$; and for S2), LA-SAGA superimposes $\bm{\lambda}_t$ to the instantaneous state (buffered resources) $\mathbf{q}_t$, and comes up with an effective multiplier as
\begin{equation}\label{eq.effect-dual}
\underbrace{~~~~\bm{\theta}_t~~~~}_{\rm effective~multiplier}=\underbrace{~~~~~\bm{\lambda}_t~~~~~}_{\rm statistical~learning}+~\underbrace{~~~\mu\mathbf{q}_t-\mathbf{b}~~~}_{\rm system~interaction}\!
\end{equation}
where $\mu$ tunes emphasis to statistical versus interactive state information, and $\mathbf{b}$ is a constant that corrects the possible bias in the steady-state - the intuition will become transparent soon.
Based on $\bm{\theta}_t$, the real-time IoT decision $\mathbf{x}_t$ is obtained by minimizing the Lagrangian over ${\cal X}(\mathbf{s}_t)$; that is,
\begin{equation}\label{eq.LA-xt}
\bm{\pi}_t(\mathbf{s}_t)=\mathbf{x}_t:=\arg\min_{\mathbf{x}\in {\cal X}(\mathbf{s}_t)}{\cal L}(\mathbf{x},\bm{\theta}_t;\mathbf{s}_t).
\end{equation}
Note that different from stochastic allocation that is solely based on the system feedback $\bm{\theta}_t=\mu\mathbf{q}_t$ \cite{neely2010}, and statistical learning that only relies on $\bm{\theta}_t=\bm{\lambda}_t$, LA-SAGA can take advantage of both through the use of effective multiplier $\bm{\theta}_t$.

To grasp how the effective multiplier accounts for S2), suppose that after sufficient learning processes, $\bm{\lambda}_t\approx \bm{\lambda}^*$, and $\mathbf{q}_t$ is large so that $\mu\mathbf{q}_t-\mathbf{b}\gg \mathbf{0}$. In this case, we have the ``shadow price'' $\bm{\theta}_t\gg \bm{\lambda}^*$, and thus $\mathbf{x}_t$ obtained through \eqref{eq.LA-xt} would ensure that $\mathbf{A}\mathbf{x}_t+\mathbf{c}_t<\mathbf{0}$ so that $\mathbf{q}_{t+1}<\mathbf{q}_t$ via \eqref{eq.prob-saopt2b}. Intuitively speaking, $\bm{\theta}_t$ will eventually oscillate around $\bm{\lambda}^*$, and thus $\mathbf{q}_t$ will oscillate around $\mathbf{b}/\mu$ - this also suggests a positive $\mathbf{b}$, otherwise it leads to a biased $\bm{\theta}_t\geq \bm{\lambda}^*$ since $\mathbf{q}_t\geq \mathbf{0}$.

Rigorous analysis demonstrate that through a proper selection of the bias $\bm{b}=\tilde{\cal O}(\sqrt{\mu})$, $\bm{\theta}_t$ will converge to the ${\cal O}(\sqrt{\mu})$-neighborhood of $\bm{\lambda}^*$ for \eqref{eq.dual-saopt}; formally, we have \cite{chen2016tsp}
\begin{subequations}\label{eq.lasdg}
\begin{align}
	       &\lim_{T\rightarrow \infty} ({1}/{T}) {\textstyle\sum_{t=1}^{T}} \mathbb{E}\left[f\left(\mathbf{x}_t; \mathbf{s}_t\right)\right]-f^* = {\cal O}(\mu)\\
	       &\lim_{T\rightarrow \infty} ({1}/{T}) {\textstyle\sum_{t=1}^{T}} \mathbb{E}\left[\|\mathbf{q}_t\|\right]=\tilde{\cal O}\left({1}/{\sqrt{\mu}}\right)
\end{align}
\end{subequations}
which asserts that LA-SAGA is ${\cal O}(\mu)$-optimal with an average queue
length $\tilde{\cal O}(1/{\sqrt{\mu}})$ - an elegant $[{\cal O}(\mu),\tilde{\cal O}(1/{\sqrt{\mu}})]$ tradeoff.
Comparing with the tradeoff $[{\cal O}(\mu),{\cal O}(1/\mu)]$ under Lyapunov optimization in \cite{neely2010}, LA-SAGA \cite{chen2016tsp,chen2017tcns} improves the performance in terms of constraint violations (queue lengths).

The idea of incorporating learning into network optimization is pioneered in \cite{huang2014}. However, the developed learning mechanism therein suffers from \emph{the curse of dimensionality}.
Targeting large-scale IoT networks, LA-SAGA can tackle settings with continuous $\mathbf{\cal S}$ and $\mathbf{\cal X}$ with possibly infinite elements, and still be amenable to efficient and scalable online operations.
The important implication of the learn-and-adapt scheme is that it can perform the optimal IoT management, with \emph{reduced resources and improved QoS}, namely, reduced queueing delay in data centers \cite{chen2017tcns}, faster virtual network function placement \cite{chen2018tmc}, and lower congestion in transportation networks, or smaller battery capacity in power grids \cite{li2017tsg}.

{As a closing remark of this subsection, note that while the problem considered in the semi-interactive setting here explicitly contains queueing-type constraints, the semi-interactive settings in fact cover a broader class of problems in IoT. For instance, throughput maximization in UAV-enabled wireless networks under trajectory constraints also belongs to the class of semi-interactive IoT settings \cite{zeng2017twc,wu2018}.}

\vspace{-0.1cm}
\subsection{Reinforcement learning for interactive IoT environments}\label{sec.saopt3}
The IoT environment considered in Section \ref{sec.saopt2} is semi-interactive in the sense that only the dynamic of $\mathbf{q}_{t+1}$ (but not $\mathbf{s}_{t+1}$) changes according to $\mathbf{x}_t$ through \eqref{eq.prob-saopt2b}.
To broaden the scope of the unified framework, this subsection introduces methods tailored for the \emph{fully interactive} setups, where the dynamic of IoT state $\mathbf{s}_t$ that can appear both in the objectives and the constraints is driven by the decision $\mathbf{x}_t$. 
{This setting captures the trajectory optimization in UAV-aided mobile communications, e.g., \cite{zeng2017twc,wu2018}, the dynamic caching with limited storage units, e.g., \cite{sadeghi2018}, and the route planning in intelligent transportation, e.g.,  \cite{zhang2018pieee}.}    

For simplicity, consider an IoT environment with a finite state space ${\cal S}$, and a finite action space ${\cal X}.$
The interaction between the operator and the IoT environment is uniquely captured by the transition probability of going from the current state $\mathbf{s}$ to the subsequent state $\mathbf{s}'$ under action $\mathbf{x}\in{\cal X}(\mathbf{s})\subseteq{\cal X}$, given by $[\mathbf{P}^{\mathbf{x}}]_{\mathbf{s}\mathbf{s}'}:=\mathbb{P}(\mathbf{s}_{t+1}=\mathbf{s}|\mathbf{s}_t=\mathbf{s}',\mathbf{x}_t=\mathbf{x})$.
Similar to \eqref{eq.prob-saopt}, the goal is to determine a possibly randomized policy $\bm{\pi}$ that given a state $\mathbf{s}_t$, generates $\mathbf{x}_t=\bm{\pi}(\mathbf{s}_t)$ so as to minimize the total discounted cost\footnote{For simplicity, the infinite horizon \emph{discounted} formulation is considered --- a slight mismatch with the generic one \eqref{eq.allprob}. Other formulations with \emph{constraints} or \emph{average} costs can be also considered with additional assumptions \cite{altman1999}.}, that is
\begin{align}\label{eq.prob-saopt3}
~\mini_{\{\bm{\pi}(\mathbf{s}_t)\in {\cal X}(\mathbf{s}_t),\,\mathbf{s}_t\}}~\lim_{T\rightarrow\infty}\mathbb{E}\left[\sum_{t=1}^T \gamma^{t-1} f\left(\bm{\pi}(\mathbf{s}_t);\mathbf{s}_t\right)\right]~
\end{align}
where $\gamma\in(0,1)$ is a discounting factor, and $\mathbb{E}$ is taken over the sample path of $\{\mathbf{s}_t\}$, as well as the random policy $\bm{\pi}$.

For a fixed policy $\bm{\pi}$, the state value function is defined as
\begin{equation}
	V_{\bm{\pi}}(\mathbf{s}):=\lim_{T\rightarrow\infty}\mathbb{E}\left[\sum_{t=1}^T \gamma^{t-1} f\left(\bm{\pi}(\mathbf{s}_t);\mathbf{s}_t\right)\Big|\mathbf{s}_1=\mathbf{s}\right]
\end{equation}
and the state-action value function (so-termed Q-function) is $Q_{\bm{\pi}}(\mathbf{s},\mathbf{x}):=f\left(\mathbf{x};\mathbf{s}\right)+\gamma \mathbb{E}_{\mathbf{s}'|\mathbf{s},\mathbf{x}}\left[V_{\bm{\pi}}(\mathbf{s}')\right]$,
where $\mathbb{E}$ is taken over the one-step transition from the current state $\mathbf{s}$ to $\mathbf{s}'$ under action $\mathbf{x}$.
With the optimal policy ${\bm{\pi}}^*$, we have that\footnote{We interchangeably use $Q^*(\mathbf{s},\mathbf{x})=Q_{\bm{\pi}^*}(\mathbf{s},\mathbf{x})$ and $V^*(\mathbf{s})=V_{\bm{\pi}^*}(\mathbf{s})$.} 
\begin{equation}
	{\bm{\pi}}^*(\mathbf{s}):=\arg\min_{\mathbf{x}\in {\cal X}(\mathbf{s})}Q^*(\mathbf{s},\mathbf{x})
\end{equation}
and $V^*(\mathbf{s})=Q^*(\mathbf{s},{\bm{\pi}}^*(\mathbf{s}))$. Furthermore, the optimality condition of \eqref{eq.prob-saopt3} that is termed Bellman optimality equation can be written as (e.g., \cite{sutton1998})
\begin{equation}\label{eq.bellman}
Q^*(\mathbf{s},\mathbf{x})\!=\!f\left(\mathbf{x};\mathbf{s}\right)+\gamma \mathbb{E}_{\mathbf{s}'|\mathbf{s},\mathbf{x}}\!\left[\min_{\mathbf{x}'\in {\cal X}(\mathbf{s})}\! Q^*(\mathbf{s}',\mathbf{x}')\right]\!,~\forall \mathbf{x},\mathbf{s}
\end{equation}
which is a system of \emph{nonlinear} equations of $\mathbf{Q}^*\in \mathbb{R}^{|{\cal S}|\times|{\cal X}|}$. 

Switching the goal from \eqref{eq.prob-saopt3} to the fixed point of the Bellman optimality equation \eqref{eq.bellman}, a classical yet popular approach is the so-termed Q-learning algorithm \cite{watkins1992}:

\noindent S1) At slot $t$, select the decision $\mathbf{x}_t$ by
\begin{equation} \label{eq.q-learning1}
	\bm{\pi}_t(\mathbf{s}_t)=\mathbf{x}_t:= \left\lbrace \begin{array}{ll}\arg \min \limits _{\mathbf{x}\in {\cal X}(\mathbf{s}_t)} {Q}_t\left(\mathbf{s}_t,\mathbf{x}\right) &\textrm {w.p. }~1-\epsilon _t \\ \textrm {random}~\mathbf{x} \in \mathcal{\cal X}(\mathbf{s}_t) & \textrm {w.p.}~~~~\epsilon _t\end{array} \!\right.\!\!
\end{equation}
where $\epsilon _t>0$ is a pre-defined exploration constant, and $\mathbf{s}_{t+1}$ is generated according to $\mathbb{P}(\mathbf{s}_{t+1}=\mathbf{s})=[\mathbf{P}^{\mathbf{x}_t}]_{\mathbf{s}_t\mathbf{s}}$.

\noindent S2) Update the state-action value function as
\begin{align}\label{eq.q-learning2}
\!\!\!Q_{t+1}(\mathbf{s}_t,\mathbf{x}_t)=\,&Q_t(\mathbf{s}_t,\mathbf{x}_t)\nonumber\\
-\,&\alpha_t\left(f\left(\mathbf{x}_t;\mathbf{s}_t\right)+\gamma\!\! \min_{\mathbf{x}\in {\cal X}(\mathbf{s}_{t+1})}\! Q_t(\mathbf{s}_{t+1},\mathbf{x})\right)
\end{align}
where $\alpha_t$ is a pre-defined stepsize. 
{Note that different from Sections \ref{sec.saopt1} and \ref{sec.saopt2}, the explicit form of the objective function $f\left(\,\cdot\,;\mathbf{s}_t\right)$ does not need to be known per slot $t$. Instead, only the functions values $\{f\left(\mathbf{x}_{\tau};\mathbf{s}_{\tau}\right)\}_{\tau=1}^t$ along the trajectory $(\mathbf{s}_1, \mathbf{x}_1),\cdots,(\mathbf{s}_t, \mathbf{x}_t)$ are assumed to be known. 
With properly selected $\{\epsilon _t,\alpha_t\}$, the simple Q-learning algorithm is provably convergent under the finite state and action spaces (a.k.a. \emph{tabular} case) \cite{sutton1998}. To date, convergence of Q-learning and its variants is mostly asserted for the tabular case.}

To scale up Q-learning in the large-scale settings, recent efforts have been devoted to infer $\mathbf{Q}$ by minimizing the residual of the Bellman optimality equation \eqref{eq.bellman}; that is,
\begin{equation}\label{eq.ls-bellman}
\min_{\mathbf{Q}}\, \sum_{\mathbf{x},\mathbf{s}}\!\left(\!Q(\mathbf{s},\mathbf{x})-f\left(\mathbf{x};\mathbf{s}\right)-\gamma \mathbb{E}_{\mathbf{s}'|\mathbf{s},\mathbf{x}}\left[\min_{\mathbf{x}'\in {\cal X}(\mathbf{s})} Q(\mathbf{s}',\mathbf{x}')\right]\!\right)^2\!\!.
\end{equation}
Albeit its simple expression, several fundamental challenges arise when solving this fitting problem \eqref{eq.ls-bellman}:

\noindent C1) the optimization scale can be prohibitively huge due to the possibly large state and action spaces;

\noindent C2) the unknown conditional expectation $\mathbb{E}_{\mathbf{s}'|\mathbf{s},\mathbf{x}}$ inside the square loss prevents an easy unbiased gradient estimator; and,

\noindent C3) the max operator inside the square loss introduces non-smoothness and non-convexity when performing optimization.

To tackle C1), function approximation methods have been studied using linear or nonlinear (random) basis functions \cite{busoniu2010,shen2018aistats}. 
Roughly speaking, given a state-action pair $(\mathbf{x},\mathbf{s})$ along with its pre-defined feature vector $\bm{\phi}_{\mathbf{x},\mathbf{s}}\in\mathbb{R}^d$, existing approaches will approximate the Q-function by $Q(\mathbf{s},\mathbf{x}):=\mathbf{z}^{\top}(\bm{\phi}_{\mathbf{x},\mathbf{s}})\bbtheta$, where $\mathbf{z}(\bm{\phi}_{\mathbf{x},\mathbf{s}})\in\mathbb{R}^{2D}$ is a lifted feature vector (e.g., random features or outputs of deep neural networks) generated from  $\bm{\phi}_{\mathbf{x},\mathbf{s}}$ and $\bbtheta\in\mathbb{R}^{2D}$ is the wanted parameter vector. To this end, the task of finding the {\small$|{\cal S}|\times |{\cal A}|$} function (matrix) $\mathbf{Q}$ reduces to finding the $2D$-dimensional vector $\bbtheta$. 
Along this line, several recent works based on primal-dual solvers have made significant progress on simultaneously resolving C1) and C2) \cite{dai2017aistats,wei2017}. 
Regarding C3), while it is still an active research area, approaches leveraging smoothing techniques for nonsmooth functions in convex optimization have shed light on promising remedies \cite{dai2017nips,nachum2017}. 

{In addition to value iteration-based methods such as Q-learning, approaches based on direct policy search such as policy gradients and actor-critic methods are also prevalent nowadays, e.g., \cite{sutton2000policy,schulman2015,zhang18icml}. 
This key idea behind policy gradient is to update the ${\bm{\theta}}$-parametrized policy ${\bm{\pi}}_{\bm{\theta}}$ using the gradient of the discounted objective \eqref{eq.prob-saopt3} with respect to the policy parameters \cite{sutton2000policy}. Convergence of the policy gradient with deep neural networks or kernel-based function approximators is now better understood than Q-learning, along with the limitations of policy gradient-based methods that arise from their high variance.} 

We conclude this section by remarking that approaches in light of the offline-aided-online learning have also been studied for \eqref{eq.prob-saopt3} under the name of \emph{experience replay}, which achieves tremendous success in various artificial intelligence tasks \cite{mnih2015}.


\section{Scalability in online learning for IoT}
\label{sec:scale}
The IoT settings considered in Section \ref{sec:adapt} involve slow-varying IoT dynamics that are (asymptotically) stationary relative to the timescale of making decisions.
In large-scale IoT however, real-time control and communications entail slow and fast time scales that prompt scalable online solvers for generally nonstationary settings --- the topics of this section.

\begin{figure}[t]
\centering
\includegraphics[width=0.46\textwidth]{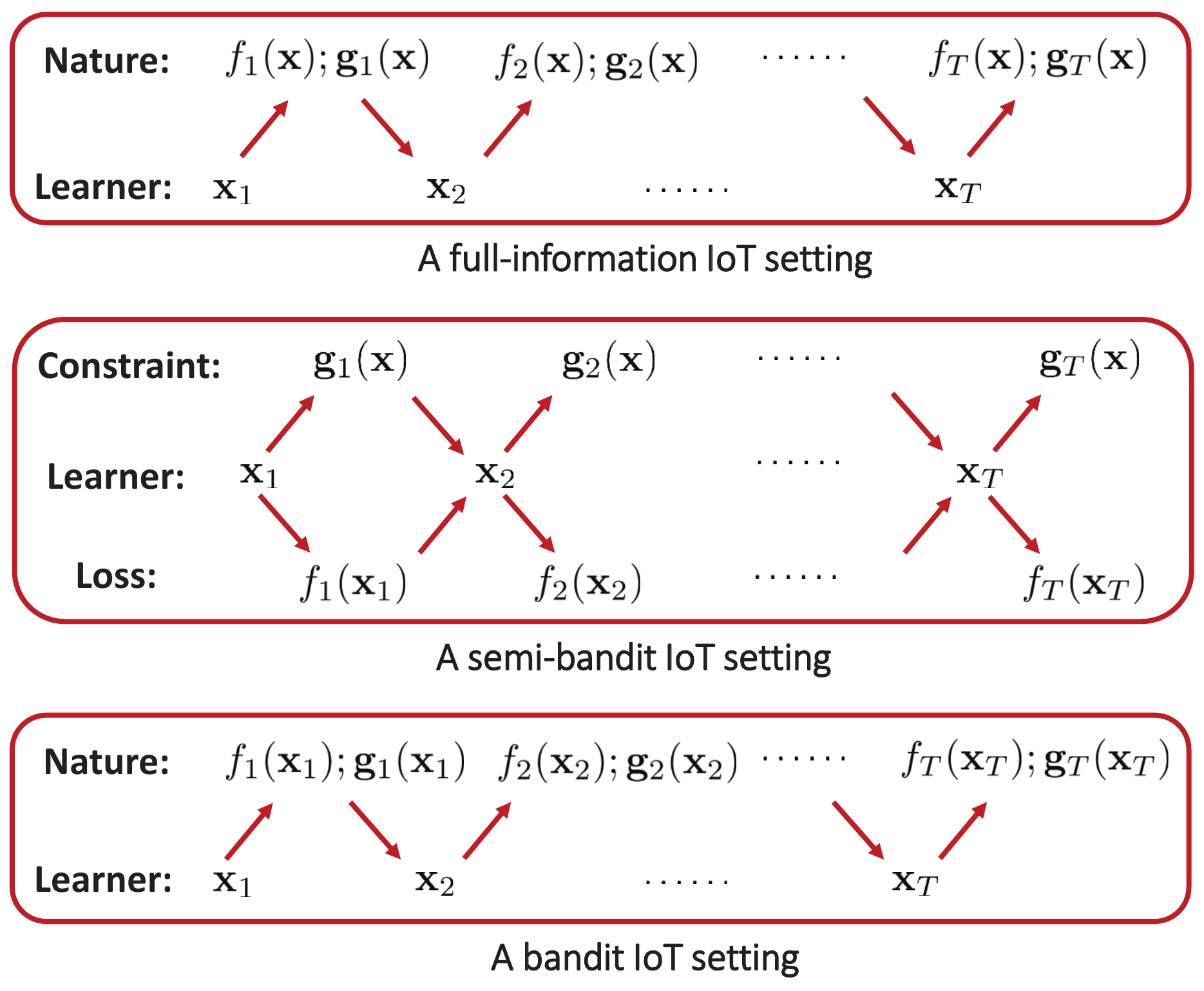}
\vspace{-0.1cm}
\caption{\small Three forms of feedback in IoT environments (termed nature) correspond to three different types of information oracle in Table \ref{tab:overview}.} \label{fig:ocoIoT-comp}
\end{figure}

In addition to the general non-stationarity, special attention will be given to approaches designed under limited information about the environment, or equivalently, solvers requiring limited computation and communication resources to sense the environment. 
Corresponding to different information that may be available in IoT, we outline three classes of scalable online learning approaches; see also Fig. \ref{fig:ocoIoT-comp} for a comparison.

\vspace{-0.2cm}

\subsection{Constrained online learning for IoT management}\label{sec:scale-1}
Consider a finite time horizon $T$. Per slot $t$, the IoT operator selects an action $\mathbf{x}_t$ from a known and fixed \emph{convex} set ${\cal X}\subseteq\mathbb{R}^d$,
and the IoT environment (a.k.a. nature in OCO) then reveals a loss $f_t: \mathbb{R}^d\rightarrow \mathbb{R}$, along with a time-varying (possibly adversarial) penalty function
$\mathbf{g}_t: \mathbb{R}^d\rightarrow \mathbb{R}^N$.
The latter leads to a time-varying constraint
$\mathbf{g}_t(\mathbf{x})\leq \mathbf{0}$, which is driven by
the unknown IoT dynamics. 
As in \eqref{eq.allprob}, the goal here is to generate a sequence
of decisions that minimize
the aggregate loss, and ensure that the constraints are satisfied in the
long term on average.
Specifically, we wish to solve
\begin{empheq}[box=\fbox]{equation}\label{eq.prob-oco}
\mini_{\{\mathbf{x}_t\in {\cal X},\forall t\}} ~\sum_{t=1}^T f_t(\mathbf{x}_t)\quad
\text{subject to}~~~\sum_{t=1}^T \mathbf{g}_t(\mathbf{x}_t) \leq \mathbf{0}.
\end{empheq}
Comparing with the generic problem \eqref{eq.allprob}, we keep the time-varying IoT state $\mathbf{s}_t$ implicit in \eqref{eq.prob-oco}, e.g., $f_t(\mathbf{x}_t) :=f(\mathbf{x}_t;\mathbf{s}_t)$ and $\mathbf{g}_t(\mathbf{x}_t) :=\mathbf{g}(\mathbf{x}_t;\mathbf{s}_t)$, since the algorithms introduced in this section may not need to directly sense the state $\mathbf{s}_t$. 
For \eqref{eq.prob-oco}, if $\{f_t,{\bf g}_t\}$ are \emph{known} and $T$ is not prohibitively large, the optimal decisions can be found using any off-the-shelf batch solver. Along with the potentially high complexity of batch solvers, a key challenge is that loss and constraint functions in dynamic IoT setups are often \emph{unknown} before allocating resources, due to unpredictable channel blocking, in millimeter wave links, due to the unpredictable routing, network congestion, device malfunctions, and nowadays malicious attacks.

Consider an edge layer with low-power sensors; a fog with $N$ nodes in ${\cal N}$; and, a cloud with multiple computing centers~\cite{chiang2016}.
Per slot $t$, each node $n$ collects data requests $d_t^n$ from nearby sensors, and has to decide among three options:

\noindent i) offloading an amount ${\chi}_t^n$ (from $d_t^n$) to the cloud;

\noindent ii) offloading  $x_t^{nm}$  to node $m$ for collaborative computing; and,

\noindent iii) processing an amount $x_t^{nn}$ using the in-situ fog servers.

\begin{figure}[t]
\vspace{-0.15cm}
\hspace{-0.15cm}
\includegraphics[width=0.49\textwidth]{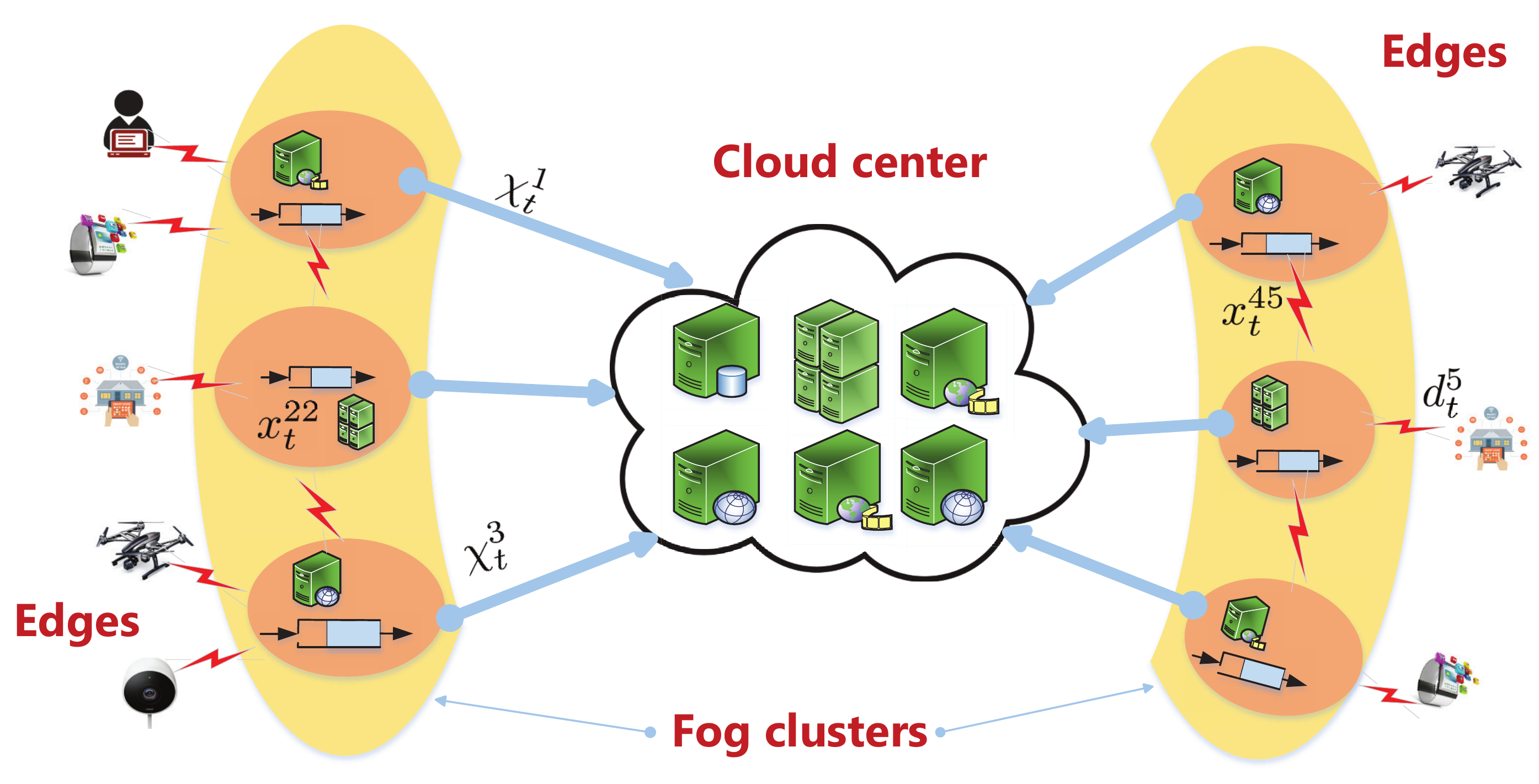}
\vspace{-0.5cm} \caption{\small A diagram for mobile computation offloading: IoT devices at the edge layer; fog clusters contain locally connected fog nodes, and the data center in the cloud layer.} \label{fig:edge-system}
\vspace{-0.2cm}
\end{figure}

\noindent Variable $\mathbf{x}_t$ consists of all the decisions in i) - iii); see Fig. \ref{fig:edge-system}.

Supposing that each fog node has a local queue to buffer unserved workloads, a long-term constraint is imposed to ensure that the cumulative amount of served workloads is no less than the arrived amount over $T$ slots; that is,
\begin{align}
&\sum_{t=1}^T g_t^n(\mathbf{x}_t)\leq\! 0,\,\forall n \nonumber \\
&g_t^n(\mathbf{x}_t):=d_t^n+\!\sum_{m\in{\cal N}_n^{\rm in}} x_t^{mn}-\!\!\sum_{m\in{\cal N}_n^{\rm out}} x_t^{nm}-{\chi}_t^n-x_t^{nn}
\label{eq.long-contrs}
\end{align}
where ${\cal N}_n^{\rm in}$ (${\cal N}_n^{\rm out}$) is the set of fog nodes with in-coming (out-going) links to (from) node $n$. Clearly, amounts ${\chi}_t^n$, $x_t^{nm}$, and $x_t^{nn}$ have caps depending on the communication protocols and computing cores in use.
With $\bar{\mathbf{x}}$ collecting all these caps, the feasible set is ${\cal X}\!:=\!\{\mathbf{0}\leq \mathbf{x}_t\leq\bar{\mathbf{x}}\}$.

Among candidate figures of merit in optimizing $\mathbf{x}_t$, is network delay of the online edge processing and offloading decisions \cite{chen2018iot,barbarossa2018edge}.
Specifically, the latency associated with ${\chi}_t^n$ is mainly due to the communication delay, which can be modeled as a time-varying convex function $l_t^n({\chi}_t^n)$.
Likewise, the communication delay related to $x_t^{nm}$ is denoted by $l_t^{nm}(x_t^{nm})$.
In addition, latency pertaining to $x_t^{nn}$ comes from  its limited computation capability, which can be modeled as a function $h_t^n(x_t^{nn})$ capturing dynamics during the computing processes.

The overall performance in allocating $\mathbf{x}_t$ is quantified by aggregate latency metrics. Those include computational ($l_t$) and communication delays ($h_t$), namely
\begin{equation}\label{eq.netcost}
\!f_t(\mathbf{x}_t)\!:=\!\!\sum_{n\in
{\cal N}}\!\!\Big(l_t^n({\chi}_t^n)+\textstyle\sum_{m\in {\cal N}_n^{\rm out}}\!l_t^{nm}(x_t^{nm})+h_t^{n}(x_t^{nn})\!\Big).\!\!\!
\end{equation}
While the aggregate delay in some cases cannot directly reflect user experience, a viable alternative is the maximum of computational and communication delays; see e.g., \cite{chen2018iot}. 
{While the average-delay objective presumed in \eqref{eq.prob-oco} may not be the optimal performance metric in some mission critical applications, our formulation can also cover the probabilistic delay requirements. The per-slot objective of the latter is an indicator function of the delay given by
\begin{equation}\label{eq.netcostinde}
\!\!f_t(\mathbf{x}_t)\!:=\sum_{n\in
{\cal N}}\!\mathds{1}\!\Bigg\{\!\Big(l_t^n({\chi}_t^n)+\!\!\!\sum_{m\in {\cal N}_n^{\rm out}}\!l_t^{nm}(x_t^{nm})+h_t^{n}(x_t^{nn})\!\Big)\!\leq\! \bar{l}\Bigg\}\!\!
\end{equation}
where $\bar{l}$ is a pre-defined upper bound of user delay.
The price paid is that the resultant problem is nonconvex, which can be tackled by e.g., the approach in Section \ref{sec:scale-3}.}

With $f_t(\mathbf{x}_t)$ as in \eqref{eq.netcost} and constraints as in \eqref{eq.long-contrs}, the solution of \eqref{eq.prob-oco} aims to minimize the aggregate delay, while serving all IoT demands in the long term. Looking forward, more intriguing is to find such an optimal strategy in a fully \emph{causal setting}, where $\{f_t(\mathbf{x}_t), d_t^n\}$ are unknown when deciding $\mathbf{x}_t$, but are revealed at the end of slot $t$ after deciding $\mathbf{x}_t$.

To gauge the performance of online decisions, static regret is adopted by OCO to measure how far the aggregate loss of an OCO algorithm is from the best
fixed solution in hindsight \cite{zinkevich2003}. Since a \textit{static regret} relies on a rather coarse benchmark, which is less useful in dynamic IoT \cite{chen2017tsp},
we are motivated to pursue the so-termed \textit{dynamic regret} given by
\begin{subequations}\label{eq.dyn-reg}
	\begin{align}
&   {\rm Reg}^{\rm d}_T:=\sum_{t=1}^T f_t(\mathbf{x}_t)-\sum_{t=1}^T f_t(\mathbf{x}_t^*)\\
    &{\rm with}~~\mathbf{x}_t^*\!\in\arg\min_{\mathbf{x}\in {\cal X}}f_t(\mathbf{x}),~~{\rm subject~to}~~\mathbf{g}_t(\mathbf{x})\!\leq\!\mathbf{0}
\end{align}
\end{subequations}
where the benchmark is now formed using the best sequence $\{\mathbf{x}_t^*\}$ for the instantaneous problem, subject to the instantaneous constraint.
The metric in (5) is more suitable for assessing performance of dynamic IoT networks than its static counterpart in  \cite{zinkevich2003}, because a sub-linear dynamic regret implies a sub-linear static one, but the converse is not true.

Regarding feasibility of online decisions, the \textit{dynamic fit} is also useful to quantify the accumulated violations, that is
\begin{align}\label{eq.dyn-fit}
    {\rm Fit}^{\rm d}_T:=\Bigg\|\Bigg[\sum_{t=1}^T \mathbf{g}_t(\mathbf{x}_t)\Bigg]^+\Bigg\|.
\end{align}
The long-term constraint implicitly assumes that the instantaneous constraint violations can be compensated by subsequent strictly feasible decisions, thus allowing adaptation of fog decisions to the unknown dynamics of IoT user demands.

Under the metrics in \eqref{eq.dyn-reg} and \eqref{eq.dyn-fit}, an ideal algorithm will be  one that achieves both sub-linear dynamic regret and sub-linear
dynamic fit. A sub-linear dynamic regret implies ``no-regret''
relative to the clairvoyant dynamic solution on the long-term
average; i.e., $\lim_{T\rightarrow \infty}{{\rm Reg}_{T}^{\rm
d}}/{T}=0$, while a sub-linear dynamic fit indicates that the online
strategy is also feasible on average; i.e., $\lim_{T\rightarrow
\infty}{{\rm Fit}_{T}^{\rm d}}/{T}=0$.

With $\bm{\lambda}\in\mathbb{R}^N_+$ denoting the Lagrange multiplier vector, the Lagrangian of \eqref{eq.prob-oco} is
\begin{align}\label{eq.Lam}
{\cal L}_t(\mathbf{x},\bm{\lambda}):=f_t(\mathbf{x})+\bm{\lambda}^{\top}\mathbf{g}_t(\mathbf{x}).
\end{align}

Building on \eqref{eq.Lam}, an online scheme termed modified saddle-point (MOSP) approach has been developed first in \cite{chen2017tsp} and later in \cite{yu2017nips}. We use the low-complexity variant in \cite{yu2017nips} for the subsequent illustration.
Given $\mathbf{x}_t$ and $\bm{\lambda}_t$, the decision
$\mathbf{x}_{t+1}$ is
\begin{equation}\label{eq.primal}
       \mathbf{x}_{t+1}={\cal P}_{\cal X}\!\left(\mathbf{x}_t-\alpha\nabla_{\mathbf{x}}{\cal L}_t(\mathbf{x}_t,\bm{\lambda}_t)\right)
\end{equation}
where ${\cal P}_{\cal X}(\mathbf{y})\!:=\!\argmin_{\mathbf{x}\in{\cal X}}\|\mathbf{x}-\mathbf{y}\|^2$; $\alpha$ is a pre-defined constant; and, $\nabla_{\mathbf{x}}
{\cal L}_t(\mathbf{x}_t,\bm{\lambda}_t)=\nabla f_t(\mathbf{x}_t)+\nabla^{\top} \mathbf{g}_t(\mathbf{x}_t)\bm{\lambda}_t$ is the gradient of ${\cal L}_t(\mathbf{x},\bm{\lambda}_t)$ with respect to (w.r.t.) $\mathbf{x}$.
In addition, the dual update takes the modified online gradient ascent form
\begin{equation}\label{eq.dual}
        \bm{\lambda}_{t+1}=\left[\bm{\lambda}_t+\mu (\mathbf{g}_t(\mathbf{x}_t)+\nabla^{\top} \mathbf{g}_t(\mathbf{x}_t)(\mathbf{x}_{t+1}-\mathbf{x}_t))\right]^{+}
\end{equation}
where $\mu$ is the stepsize, and $\mathbf{g}_t(\mathbf{x}_t)$ the gradient of ${\cal L}_t(\mathbf{x}_t,\bm{\lambda})$ w.r.t. $\bm{\lambda}$.
Note that \eqref{eq.dual} is a modified gradient update since the dual variable is updated along the first-order approximation of $\mathbf{g}_t(\mathbf{x}_{t+1})$ at $\mathbf{x}_t$ rather than the commonly used $\mathbf{g}_t(\mathbf{x}_t)$.

With properly chosen stepsizes, MOSP enjoys dynamic regret and fit bounded by~\cite{chen2017tsp}
\begin{align}\label{sec4-them1}
 	  {\rm Reg}^{\rm d}_T\!=\! {\cal O}\Big(\mathbb{V}(\mathbf{x}_{1:T}^*)T^{\frac{1}{2}}\Big){\rm~~and~~}{\rm Fit}^{\rm d}_T
	={\cal O}\big(T^{\frac{1}{2}}\big)
\end{align}
where $\mathbb{V}(\mathbf{x}_{1:T}^*)$ is the accumulated
variation of the per-slot minimizers $\mathbf{x}^*_t$ in \eqref{eq.dyn-reg} given by $ \mathbb{V}(\mathbf{x}_{1:T}^*):=\sum_{t=1}^T \|\mathbf{x}^*_t-\mathbf{x}^*_{t-1}\|$.
In words, MOSP's dynamic fit is sub-linear, and its dynamic regret is also sub-linear, so long as the variation of the minimizers is slow enough; i.e., $\mathbb{V}(\mathbf{x}_{1:T}^*)=\mathbf{o}(\sqrt{T})$.

Relevant approaches developed in similar settings also include \cite{li2018acc,bernstein2018,dall2018tsg}. 
Specifically, OCO with switching cost has been studied in \cite{li2018acc}, and feedback-based tracking algorithms have been developed in \cite{bernstein2018,dall2018tsg}. 

\begin{remark}[Learning via task-adaptive stepsizes]
The primal update \eqref{eq.primal} can be refined by adjusting each entry of the gradient using a \emph{per-entry stepsize} in accordance with ``each thing'' in IoT applications \cite{chen2018iot}.
Such an adaptive stepsize can be regarded as an inexpensive approximation of the Hessian used in the online Newton iteration \cite{duchi2011jmlr}.
Using edge computing as a paradigm, \cite{chen2018iot} showed that task-adaptive stepsizes can markedly reduce the network delay when the underlying IoT tasks are heterogeneous, where the resultant gradients could have \emph{distinct} orders of magnitude over different coordinates.
\end{remark}

\vspace{-0.2cm}

\subsection{Constrained convex bandit learning for IoT management}\label{sec:scale-2}
The online recursions \eqref{eq.primal} and \eqref{eq.dual} remain operational under the premise that the loss functions are known, or, their gradients are readily available.
Clearly, none of these assumptions is always satisfied in IoT, because i) the loss function capturing user dissatisfaction, e.g., service latency or reliability, can be hard to model in dynamic settings; and, ii) even if modeling is possible, the low-power devices may not afford the complexity of running statistical learning tools such as deep neural networks online. These considerations motivate online \emph{bandit} saddle-point (BanSP) methods to broaden the scope of MOSP to IoT settings where the gradient is unavailable or computationally costly \cite{chen2017iot}.

The key idea behind bandit learning is to construct (preferably stochastic) gradient estimates using limited \emph{function value} information \cite{flaxman2005,nesterov2017}.
Consider first a learner only observing the value of $f_t(\mathbf{x})$ at a single point $\mathbf{x}$ per slot $t$. The crux is to construct a (possibly unbiased) estimate of the gradient using this single piece of feedback - what is interestingly possible by one \emph{random} function evaluation \cite{flaxman2005}. The intuition is easy to grasp in the one-dimensional case: For a binary variable $u$ taking values $\{-1,1\}$ equiprobably, and a small $\delta>0$, the difference approximation of the derivative $f_t'$ at $x$ yields
\begin{equation}
f_t'(x)\approx \frac{f_t(x+\delta )-f_t(x-\delta)}{2\delta}=\mathbb{E}_{u}\left[\frac{u}{\delta}f_t(x+\delta u)\right]
\end{equation}
where the equality follows from the definition of expectation. Dropping $\mathbb{E}_{u}$, the scaled single-value evaluation $f_t(x+\delta u)u/\delta$ is a nearly unbiased estimator of $f_t'(x)$. Generalizing this approximation to higher dimensions, with a random vector $\mathbf{u}$ drawn from the surface of a unit sphere, the scaled function evaluation at a perturbed point $\mathbf{x}+\delta\mathbf{u}$ yields an estimate of the gradient $\nabla f_t(\mathbf{x})$, given by \cite{flaxman2005}
\begin{equation}\label{eq.grad1}
\nabla f_t(\mathbf{x})\approx \mathbb{E}_{\mathbf{u}}\left[\frac{d}{\delta}f_t(\mathbf{x}+\delta \mathbf{u})\mathbf{u}\right]:=\mathbb{E}_{\mathbf{u}}\left[\hat{\nabla} f_t(\mathbf{x})\right]
\end{equation}
where we define one-point gradient $\hat{\nabla} f_t(\mathbf{x}):=\frac{d}{\delta}f_t(\mathbf{x}+\delta \mathbf{u})\mathbf{u}$.

Building upon \eqref{eq.grad1}, consider the primal update (cf. \eqref{eq.primal})
\begin{equation}\label{eq.primal2}
       \hat{\mathbf{x}}_{t+1}={\cal P}_{(1-\gamma){\cal X}}\!\left(\hat{\mathbf{x}}_t-\alpha\left(\hat{\nabla} f_t(\hat{\mathbf{x}}_t)+\nabla^{\top} \mathbf{g}_t(\hat{\mathbf{x}}_t)\bm{\lambda}_t\right)\right)
\end{equation}
where $(1-\gamma){\cal X}:=\{(1-\gamma)\mathbf{x}:\mathbf{x}\in{\cal X}\}$ is a subset of ${\cal X}$,  and $\gamma\in[0,1)$ is a pre-selected constant dependent on $\delta$.
In the full-information case, $\mathbf{x}_t$ in \eqref{eq.primal} is the learner's action, whereas in the bandit case the learner's action is $\mathbf{x}_t:=\hat{\mathbf{x}}_t+\delta \mathbf{u}_t$, which is the point for function evaluation instead of $\hat{\mathbf{x}}_t$ in \eqref{eq.primal2}. Projection in \eqref{eq.primal2} is on a smaller convex set $(1-\gamma){\cal X}$ in \eqref{eq.primal2}, which ensures feasibility of the perturbed $\mathbf{x}_t\in {\cal X}$. Similar to \eqref{eq.dual}, the dual update of BanSP is given by
\begin{equation}\label{eq.dual2}
        \bm{\lambda}_{t+1}=\left[\bm{\lambda}_t+\mu (\mathbf{g}_t(\hat{\mathbf{x}}_t)+\nabla^{\top} \mathbf{g}_t(\hat{\mathbf{x}}_t)(\hat{\mathbf{x}}_{t+1}-\hat{\mathbf{x}}_t))\right]^{+}
\end{equation}
where $\hat{\mathbf{x}}_t$ rather than $\mathbf{x}_t$ is used in this update.
Compared with \eqref{eq.primal}-\eqref{eq.dual}, the updates \eqref{eq.primal2}-\eqref{eq.dual2} with one-point bandit feedback do not increase computation or memory requirements; hence, they provide a light-weight surrogate for MOSP to enable gradient-free online bandit IoT network optimization.

If the mild conditions in \cite{chen2017iot} are satisfied, the online decisions generated by BanSP yield
\begin{align}\label{Them.dyn-reg1}
\!\!\!\mathds{E}\left[{\rm Reg}^{\rm d}_T\right]\!=\! {\cal O}\Big(\mathbb{V}(\mathbf{x}_{1:T}^*)T^{\frac{3}{4}}\Big){\rm ~~and~~}  \mathds{E}\left[{\rm Fit}^{\rm d}_T\right]\!=\!{\cal O}\big(T^{\frac{3}{4}}\big)
\end{align}
where $\mathbb{E}$ is taken over the sequence of the random actions $\mathbf{x}_t$ with randomness induced by $\{\mathbf{u}_t\}$ perturbations.

Depending on the underlying dynamics, BanSP can afford one or multiple loss function
evaluations (bandit feedback) per slot.
If BanSP is endowed with $M>2$ function evaluations, the gradient estimate will be more accurate by querying the function values over $M$ points in the neighborhood of $\hat{\mathbf{x}}_t$. Intuitively, the performance of BanSP will improve if multiple evaluations are available per slot. Indeed, the dynamic regret is provably ${\cal O}\big(\mathbb{V}(\mathbf{x}_{1:T}^*)T^{\frac{1}{2}}\big)$, and the dynamic fit ${\rm Fit}^{\rm d}_T 	={\cal O}\big(T^{\frac{1}{2}}\big)$~\cite{chen2017iot}, which markedly improve upon their single-point counterparts, and reduce to MOSP bounds in the full-information case (cf. \eqref{sec4-them1}).

\vspace{-0.2cm}
\subsection{Constrained multi-armed bandit learning}\label{sec:scale-3}
The salient assumption so far is that IoT decisions belong to a time-invariant convex set ${\cal X}$. However, IoT devices usually exhibit time-varying connectivity to the backbone due to mobility and cyber attacks, while network configurations are often selected from pre-determined protocols. In this context, multi-armed bandit (MAB) methods can be employed to extend BanSP when ${\cal X}$ is \emph{time-varying} and \emph{discrete} \cite{li2018uai,li2018asi}.

Consider the discrete feasible set ${\cal X}:=\{\mathbf{x}^1,\ldots,\mathbf{x}^K\}$ with total $K$ possible actions (a.k.a. arms in MAB). To account for dynamics, only the actions in ${\cal X}_t \subseteq {\cal X}$ are available per slot $t$; e.g., $\mathbf{x}_t \in {\cal X}_t$. The availability of actions could be stochastic, following a certain probability distribution; or even adversarial, in which case nature can arbitrarily choose ${\cal X}_t$.

Per slot $t$, collect the objective values of all actions into vector $\mathbf{f}_t:=[f_t(\mathbf{x}^1),\ldots,f_t(\mathbf{x}^K)]^{\top}$, and likewise the constraints into matrix $\mathbf{G}_t\!:=\![\mathbf{g}_t(\mathbf{x}^1),\ldots,\mathbf{g}_t(\mathbf{x}^K)]\!\in\!\mathds{R}^{N\times K}$.
If the learner's strategy is to select an action $\mathbf{x}_t=\mathbf{x}^k$ with $k$ from a distribution $k\sim \mathbf{p}_t\in\mathds{R}^K$, then \eqref{eq.prob-oco} can be re-formulated as an optimization problem over distributions $\{\mathbf{p}_t\}$, namely
\begin{align}\label{eq.prob2}
		\mini_{\{\mathbf{p}_t\in{\Delta({\cal X}_t)},\forall t\}}\,\sum_{t=1}^{T} \mathbf{f}_t^\top \mathbf{p}_t~~\st~~\sum_{t=1}^{T} \mathbf{G}_t\mathbf{p}_t\leq \mathbf{0}
\end{align}
where the ${\cal X}_t$-supported ``probability simplex'' is defined as
\begin{equation}\label{eq.simplex}
\Delta({\cal X}_t)\!:=\!\Bigg\{\sum_{\mathbf{x}^k\in {\cal X}_t}\! p(\mathbf{x}^k)=1;\, p(\mathbf{x}^k)\geq 0;\, p(\mathbf{x}^k)=0, \mathbf{x}^k\notin {\cal X}_t\!\Bigg\}.\!\!
\end{equation}
It is worth mentioning that $f_t(\mathbf{x}^k)$ and $\mathbf{g}_t(\mathbf{x}^k)$ are well defined even when the action $\mathbf{x}^k\notin {\cal X}_t$ is not available, and the
values $f_t(\mathbf{x}^k)$ and $\mathbf{g}_t(\mathbf{x}^k)$ are not revealed.

\begin{figure}[t]
	\centering
\vspace{-0.2cm}
	\includegraphics[width=0.47\textwidth]{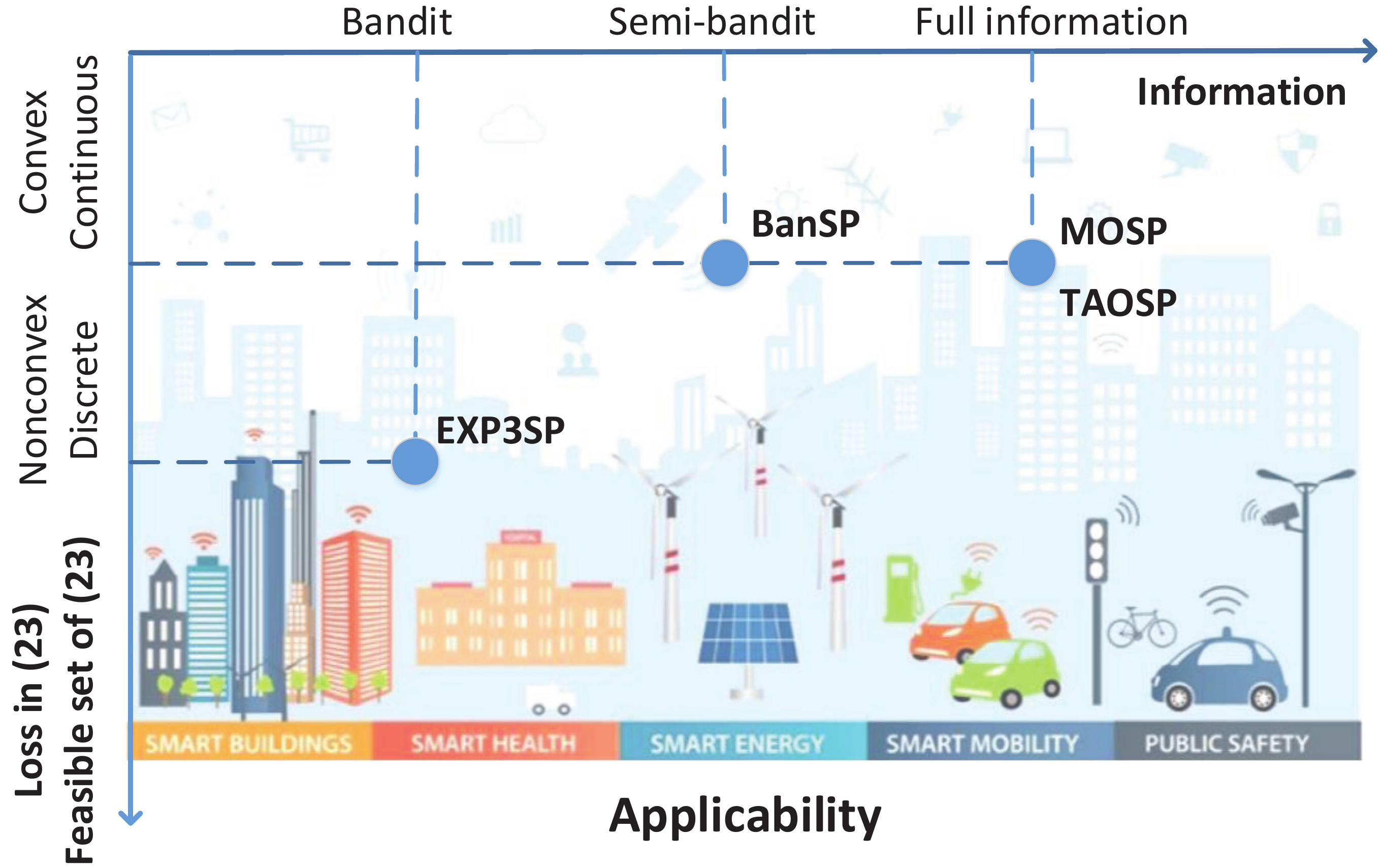}
	\caption{\small {A summary of learning approaches corresponding to Fig. \ref{fig:ocoIoT-comp}.} 
	} \label{fig:oco-IoT}
	\vspace{-0.1cm}
\end{figure}

In order to employ an MOSP solver for \eqref{eq.prob2}, the gradient of the associated Lagrangian is needed, meaning $\mathbf{f}_t$ and $\mathbf{G}_t$ must be known. The challenge is that such information is hardly available in large-scale IoT settings, where one only knows $f_t(\mathbf{x}_t)$ and $\mathbf{g}_t(\mathbf{x}_t)$ given that $\mathbf{x}_t$ is implemented.
The time-varying action set also prevents a direct implementation of BanSP to solve \eqref{eq.prob2}.
To tackle such a challenging setting, a novel EXP3SP algorithm was developed in our recent work \cite{li2018uai} that builds on the elegant \emph{exponential-weight algorithm for exploration and exploitation} (EXP3) \cite{auer2002a}.

Per slot $t$, the learner observes the action set ${\cal X}_t$, and selects $\mathbf{x}_t$ according to the current distribution $\mathbf{p}_t$ given by
\begin{equation}\label{eq.pt}
	p_t(\mathbf{x}^k) = \frac{\tilde{p}_t(\mathbf{x}^k) \mathds{1} (\mathbf{x}^k \in {\cal X}_t)}{\sum_{\mathbf{x}^k \in {\cal X}} \tilde{p}_t(\mathbf{x}^k) \mathds{1} (\mathbf{x}^k \in {\cal X}_t)},\, \forall \mathbf{x}^k \in {\cal X}
\end{equation}
where $\tilde{p}_t(\mathbf{x}^k)$ is the unnormalized weight of $\mathbf{x}^k$ at slot $t$, the value of which will be specified later. Once $f_t(\mathbf{x}_t)$ and $\mathbf{g}_t(\mathbf{x}_t)$ become available, unbiased estimates of $\mathbf{f}_t$ and $\mathbf{G}_t$ are \cite{li2018uai}
\begin{subequations}\label{eq.hatgd}
\begin{align}
\hat{f}_t(\mathbf{x}^k) &= \frac{f_t(\mathbf{x}^k) \mathds{1}(\mathbf{x}_t=\mathbf{x}^k)}{p_t(\mathbf{x}^k)},~~~\forall \mathbf{x}^k \in {\cal X}\label{eq.tildec}\\
\hat{\mathbf{g}}_t(\mathbf{x}^k) &= \frac{\mathbf{g}_t(\mathbf{x}^k) \mathds{1}(\mathbf{x}_t=\mathbf{x}^k)}{p_t(\mathbf{x}^k)},~~~\forall \mathbf{x}^k \in {\cal X}.\label{eq.tildef}
\end{align}
\end{subequations}

Adopting the gradient estimators in \eqref{eq.hatgd}, the primal update uses the exponential gradient recursion, namely, $\forall \mathbf{x}^k \in {\cal X}$
\begin{equation}\label{eq.primalupdate2}
\tilde{p}_{t+1}(\mathbf{x}^k) = \tilde{p}_t(\mathbf{x}^k)\exp\left[-\mu \Big(\hat{f}_t(\mathbf{x}^k)+\bm{\lambda}_t^{\top} \hat{\mathbf{g}}_t(\mathbf{x}^k) \Big)\right].
\end{equation}
The weight $\tilde{\mathbf{p}}_{t+1}$ is in turn used to generate the action distribution in the next slot (cf. \eqref{eq.pt}).
The dual update is
\begin{equation}\label{eq.dualupdate2}
\bm{\lambda}_{t+1} = \Big[ \bm{\lambda}_t + \mu \big(\hat{\mathbf{G}}_t\mathbf{p}_t-\delta\mu\bm{\lambda}_t\big) \Big]^+
\end{equation}
where $\delta$ is a tuned constant to ensure a bounded multiplier.

If ${\cal X}_t$ is stochastic, EXP3SP achieves both sub-linear regret and fit \cite{li2018uai}. A robust modification of EXP3SP has been also developed recently to cope with adversaries blocking access of IoT devices to their edge servers \cite{li2018asi}, while further securing edge computing and ensuring sub-linear regret and fit. A remark is now in order on a scalable rendition of EXP3SP.


\noindent
Our scalable online learning schemes are recapped in Fig. \ref{fig:oco-IoT}.



\vspace{-0.2cm}
\section{Lessons learned and the road ahead}
\label{sec:conclusion}

We have presented a unified framework for deriving and analyzing adaptive and scalable network design and resource allocation schemes for IoT. Leveraging the contemporary communication, networking and optimization advances, the resultant online learning and management policies not only facilitate low-complexity and scalable implementations with limited feedback, but also enjoy efficient adaptation to changing environments with analytical performance guarantees. 

\begin{figure}
\vspace{-1mm}
	\centering
	\includegraphics[scale=0.26]{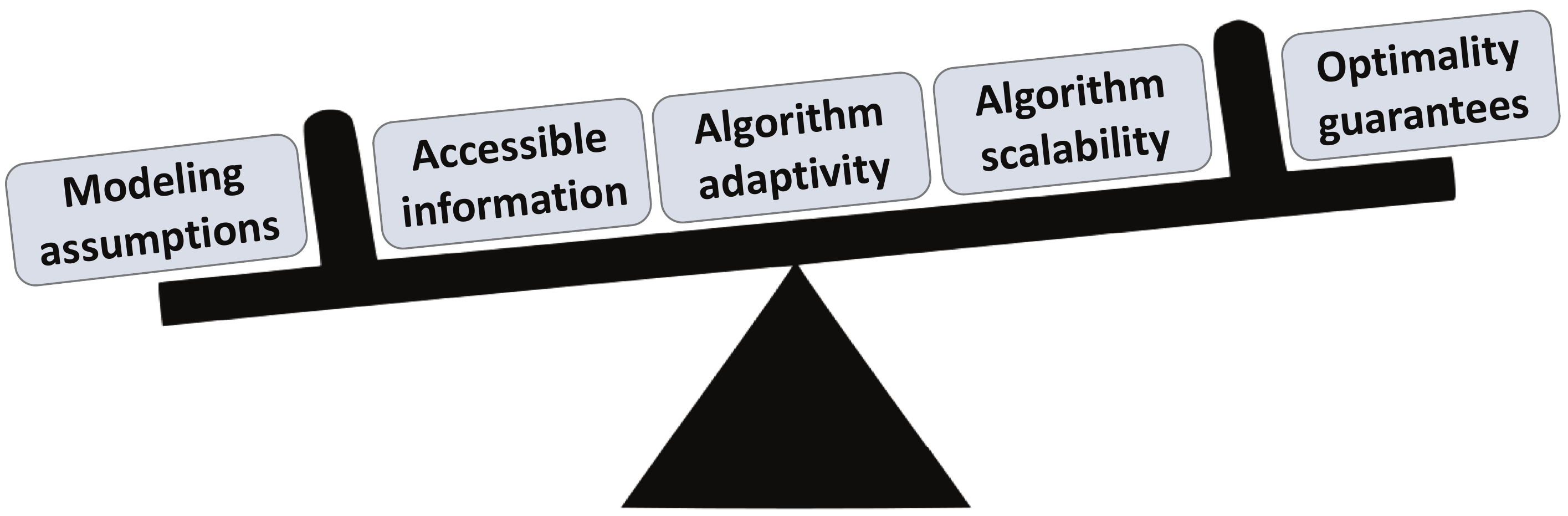}
	\caption{\small {The tradeoff among modeling assumptions, accessible information, algorithm adaptivity, scalability and optimality guarantees.}}
	\label{fig:mod-opt}
\end{figure}

The proposed framework lays a solid analytical foundation to delineate the tradeoffs among performance guarantees, degree of (non-)stationarity in modeling IoT dynamics, algorithm scalability, and levels of accessible information; see Fig. \ref{fig:mod-opt}.

\noindent\textbf{Modeling assumptions vs optimality guarantees.} 
While both deal with IoT management with \emph{unknown dynamics}, the modeling assumptions in Sections \ref{sec:adapt} and \ref{sec:scale} differ considerably. 
Specifically, those in Section \ref{sec:adapt} assume a generally \emph{stationary} IoT environment that corresponds to either the simplest i.i.d. case, or to the Markovian case  eventually converging to a stationary distribution. 
In contrast, the approaches in Section \ref{sec:scale} can afford \emph{arbitrary} dynamics even those manipulated by adversaries. 
However, such minimal assumption does not come for free. 
As a matter of fact, the performance guarantee in terms of the sublinear regret in Section \ref{sec:scale} is weaker than the optimality gap in Section \ref{sec:adapt} --- see an analytical comparison in \cite{chen2017tsp}. 
Nevertheless, as the effectiveness of the optimal solution in Section \ref{sec:adapt} also depends on the discrepancy between the real IoT settings and the modeled stationary ones, the actual online performance of these approaches requires further evaluation.   

The vantage point of this overview opens up a number
of exciting directions for future research.

\vspace{0.1cm}
{
\noindent\textbf{Distributed machine learning.} 
Considering the massive amount of mobile devices in IoT, centralized learning becomes computationally intractable, and also rises serious privacy concerns. To date, the widespread consensus is that besides data centers at the cloud, future machine learning and artificial intelligence tasks have to be performed starting from the network edge, namely mobile devices. This is the overarching goal of the emerging \emph{federated learning} paradigm \cite{mcmahan2017blog,mcmahan2017}.
Towards this goal, future challenges and opportunities include reducing the communication overhead during the distributed learning processes, and enhancing the robustness of learning algorithms under adversarial attacks. 
Recent advances in the direction of communication-efficient learning include the adaptive communication mechanism in\cite{chen2018nips} that enjoys the first provably bound on the reduced number of communication rounds. Challenges of distributed learning also lie in asynchrony and delay introduced by e.g., IoT mobility and heterogeneity. Asynchronous parallel learning schemes are thus worth investigating by leveraging advances in static optimization settings \cite{peng2016,cannelli2016}.
From distributed machine learning to distributed control, multi-agent reinforcement learning will play a critical role in distributed control for IoT \cite{lowe2017nips}.
A decentralized actor-critic algorithm has been recently developed in \cite{zhang18icml} for multi-agent reinforcement learning over networked agents, and further generalized to tasks with large continuous state and action 
spaces~\cite{zhang18cdc}.}


\vspace{0.1cm}
{
\noindent\textbf{Communication, computation and control co-design.} 
The past decade has witnessed the convergence of the communication
and computing processes \cite{barbarossa2014communicating}. The current brief is that next-generation communication networks should support emerging large-scale control applications in IoT with millions of diverse devices over a large geographical area. This calls for co-designing communication, computing, and control mechanisms. The challenges naturally arise in developing the desired network architecture, the role of different network entities, pertinent performance metrics, and the corresponding policies to simultaneously satisfy the timeliness, reliability and efficiency of all three intertwined systems.}

Over the decades, the focus of wireless communications has been anytime, anywhere, anyone connection of the humans, whereas the emerging IoT paradigm largely extends the scope of wireless networking to connecting everything, along the humans-to-things and things-to-things continuum. 
The IoT challenges such as extreme heterogeneity, unpredictable dynamics and massive scale, call for game-changing innovations in network design and management. 
We hope that the proposed unified framework can serve as a stepping stone that leads to systematic designs and rigorous analysis of adaptive and scalable learning and management schemes for IoT, and a host of new research venues to pursue. 


\balance

\end{document}